\documentclass[journal]{IEEEtran}

\usepackage{scalerel}
\usepackage{tikz}
\usetikzlibrary{svg.path}

\definecolor{orcidlogocol}{HTML}{A6CE39}
\tikzset{
  orcidlogo/.pic={
    \fill[orcidlogocol] svg{M256,128c0,70.7-57.3,128-128,128C57.3,256,0,198.7,0,128C0,57.3,57.3,0,128,0C198.7,0,256,57.3,256,128z};
    \fill[white] svg{M86.3,186.2H70.9V79.1h15.4v48.4V186.2z}
                 svg{M108.9,79.1h41.6c39.6,0,57,28.3,57,53.6c0,27.5-21.5,53.6-56.8,53.6h-41.8V79.1z M124.3,172.4h24.5c34.9,0,42.9-26.5,42.9-39.7c0-21.5-13.7-39.7-43.7-39.7h-23.7V172.4z}
                 svg{M88.7,56.8c0,5.5-4.5,10.1-10.1,10.1c-5.6,0-10.1-4.6-10.1-10.1c0-5.6,4.5-10.1,10.1-10.1C84.2,46.7,88.7,51.3,88.7,56.8z};
  }
}

\newcommand\orcidicon[1]{\href{https://orcid.org/#1}{\mbox{\scalerel*{
\begin{tikzpicture}[yscale=-1,transform shape]
\pic{orcidlogo};
\end{tikzpicture}
}{|}}}}

\usepackage{cite}
\usepackage{hyperref}
    \hypersetup{%
        pdfborder={0 0 0},
    }

\ifCLASSINFOpdf
\else

\fi

\usepackage{multirow}
\usepackage{svg}
\usepackage{lscape}
\usepackage{xcolor}
\usepackage{threeparttable}
\usepackage{pifont}

\newcommand{\cmark}{\textcolor{green}{\ding{51}}}
\newcommand{\xmark}{\textcolor{red}{\ding{55}}}

\usepackage{subcaption}
\usepackage{graphicx}
\usepackage{array}
\usepackage{booktabs}
\newcolumntype{N}{>{\centering\arraybackslash}m{.5in}}
\newcolumntype{G}{>{\centering\arraybackslash}m{2in}}

\usepackage{siunitx}
\usepackage{float}

\usepackage{amssymb}

\usepackage{acronym}
    \acrodef{ai}[AI]{Artificial Intelligence}
    \acrodef{alu}[ALU]{Arithmetic Logic Unit}
    \acrodef{api}[API]{Application Programming Interface}
    \acrodef{asic}[ASIC]{Application-Specific Integrated Circuit}
    \acrodef{bram}[BRAM]{Block Random Access Memory}
    \acrodefplural{bram}[BRAMs]{Block Random Access Memories}
    \acrodef{cgra}[CGRA]{Coarse-Grained Reconfigurable Architecture}
    \acrodef{cmn}[CMN]{Configuration Memory Node}
    \acrodefplural{cnn}[CNNs]{Convolutional Neural Networks}
    \acrodef{cpu}[CPU]{Central Processing Unit}
    \acrodef{csr}[CSR]{Control and Status Register}
    \acrodef{ddr}[DDR]{Double Data Rate}
    \acrodef{dnn}[DNN]{Deep Neural Network}
    \acrodef{dfg}[DFG]{Data-Flow Graph}
    \acrodef{dpr}[DPR]{Dynamic and Partial Reconfiguration}
    \acrodef{dma}[DMA]{Direct Memory Access}
    \acrodef{dsp}[DSP]{Digital Signal Processor}
    \acrodef{dlp}[DLP]{Data-Level Parallelism}
    \acrodef{eda}[EDA]{Electronic Design Automation}
    \acrodef{ff}[FF]{Flip Flop}
    \acrodef{fpga}[FPGA]{Field-Programmable Gate Array}
    \acrodef{fpu}[FPU]{Floating-Point Unit}
    \acrodef{fsm}[FSM]{Finite State Machine}
    \acrodef{fu}[FU]{Functional Unit}
    \acrodef{fll}[FFL]{Frequency Lock Loop}
    \acrodef{fifo}[FIFO]{First In First Out}
    \acrodef{gpu}[GPU]{Graphic Processing Unit}
    \acrodef{hls}[HLS]{High-Level Synthesis}
    \acrodef{hpc}[HPC]{High-Performance Computing}
    \acrodef{imn}[IMN]{Input Memory Node}
    \acrodef{ir}[IR]{Intermediate Representation}
    \acrodef{isa}[ISA]{Instruction Set Architecture}
    \acrodef{iot}[IoT]{Internet of Things}
    \acrodef{ii}[II]{Initiation Interval}
    \acrodef{ip}[IP]{Intellectual Propierty}
    \acrodefplural{ip}[IPs]{Intellectual Propierties}
    \acrodef{ilp}[ILP]{Instruction-Level Parallelism}
    \acrodef{llvm}[LLVM]{Low Level Virtual Machine}
    \acrodef{lut}[LUT]{Look-Up Table}
    \acrodef{mcu}[MCU]{Micro-Controller Unit}
    \acrodef{mimd}[MIMD]{Multiple Instructions Multiple Data}
    \acrodef{mmio}[MMIO]{Memory-Mapped Input/Output}
    \acrodef{noc}[NoC]{Network on Chip}
    \acrodefplural{noc}[NoCs]{Networks on Chip}
    \acrodef{nn}[NN]{Neural Network}
    \acrodef{omn}[OMN]{Output Memory Node}
    \acrodef{ooo}[OOO]{Out-of-Order}
    \acrodef{pe}[PE]{Processing Element}
    \acrodef{rocc}[RoCC]{Rocket Custom Coprocessor}
    \acrodef{simd}[SIMD]{Single Instruction Multiple Data}
    \acrodef{soc}[SoC]{System on Chip}
    \acrodefplural{soc}[SoCs]{Systems on Chip}
    \acrodef{sram}[SRAM]{Static Random Access Memory}
    \acrodef{strela}[STRELA]{STReaming-ELAstic}
    \acrodef{vlsi}[VLSI]{Very Large-Scale Integration}
    \acrodef{vpu}[VPU]{Vector Processing Unit}
    \acrodef{xheep}[X-HEEP]{eXtendable Heterogeneous Energy-Efficient Platform}

\begin{document}

\title{STRELA: STReaming ELAstic CGRA Accelerator for Embedded Systems}

\author{Daniel~V\'azquez\textsuperscript{\orcidicon{0000-0002-8617-5973}},
        Jos\'e~Miranda\textsuperscript{\orcidicon{0000-0002-7275-4616}},
        Alfonso~Rodr\'iguez\textsuperscript{\orcidicon{0000-0001-6326-743X}},
        Andr\'es~Otero\textsuperscript{\orcidicon{0000-0003-4995-7009}},
        Pasquale~Davide~Schiavone\textsuperscript{\orcidicon{0000-0003-2931-0435}},
        David~Atienza\textsuperscript{\orcidicon{0000-0001-9536-4947}}%
\thanks{This work has been supported by the project \mbox{\textit{AIQ-Ready}} \mbox{PCI2022-135077-2}, funded by the Spanish Government \textit{\mbox{MICIU/AEI} /\mbox{10.13039/501100011033}} and the European Union \textit{NextGenerationEU/PRTR}.}%
\thanks{Daniel V\'azquez (corresponding author), Alfonso Rodr\'iguez, and Andr\'es Otero are with the Centro de Electr\'onica Industrial, Universidad Polit\'ecnica de Madrid (UPM), Madrid, Spain (email: \mbox{daniel.vazquez@upm.es}; \mbox{alfonso.rodriguezm@upm.es}; \mbox{joseandres.otero@upm.es}).}%
\thanks{Jos\'e Miranda, Pascuale Davide Schiavone, and David Atienza are with the Embedded Systems Laboratory, \'Ecole Polytechnique F\'ed\'erale de Lausanne (EPFL), 1015 Lausanne, Switzerland (email: \mbox{jose.mirandacalero@epfl.ch}; \mbox{davide.schiavone@epfl.ch}; \mbox{david.atienza@epfl.ch}).}}

\markboth{V\'azquez \MakeLowercase{\textit{et al.}}: Streaming Elastic CGRA}{V\'azquez \MakeLowercase{\textit{et al.}}: Streaming Elastic CGRA}

\maketitle

\begin{abstract}
Reconfigurable computing offers a good balance between flexibility and energy efficiency. When combined with software-programmable devices such as \acp{cpu}, it is possible to obtain higher performance by spatially distributing the parallelizable sections of an application throughout the reconfigurable device, while the \ac{cpu} is in charge of control-intensive sections. This work aims to introduce an elastic \ac{cgra} integrated into an energy-efficient \mbox{RISC-V}-based \ac{soc} designed for the embedded domain. The microarchitecture of \ac{cgra} supports conditionals and irregular loops, making it adaptable to domain-specific applications. Additionally, we propose specific mapping strategies that enable the efficient utilization of the \ac{cgra} for both simple applications, where the fabric is only reconfigured once (one-shot kernel), and more complex ones, where it is necessary to reconfigure the \ac{cgra} multiple times to complete them (multi-shot kernels). Large kernels also benefit from the independent memory nodes incorporated to streamline data accesses. Due to the integration of \ac{cgra} as an accelerator of the \mbox{RISC-V} processor, a versatile and efficient framework is achieved, providing adaptability, processing capacity, and overall performance across a wide range of applications.

The design has been implemented in TSMC 65 nm technology, achieving a maximum frequency of 250 MHz. It achieves a peak performance of 1.22 GOPs computing one-shot kernels and 1.17 GOPs computing multi-shot kernels. The best energy efficiency is 72.68 MOPs/mW for one-shot kernels and 115.96 MOPs/mW for multi-shot kernels. The design integrates power and clock-gating techniques to tailor the architecture to the embedded domain while maintaining performance. The best speed-ups are 17.63x and 18.61x for one-shot and multi-shot kernels. The best energy savings in the \ac{soc} are 9.05x and 11.10x for one-shot and multi-shot kernels.
\end{abstract}

\begin{IEEEkeywords}
Reconfigurable Computing, Embedded Systems, \ac{cgra}, Low Power, \mbox{RISC-V}.
\end{IEEEkeywords}

\IEEEpeerreviewmaketitle

\section{Introduction}

\IEEEPARstart{T}{he} end of Moore's law is forcing the industry to explore alternatives beyond the miniaturization of the existing microelectronic devices to continue supporting the increasing computation demands in embedded systems. Among them is circuit specialization, which involves designing hardware optimized for specific computations as an alternative to general-purpose computing. Specialized circuits can perform their intended tasks more efficiently than general-purpose \acp{cpu}, featured with programmable fixed circuitry via software. Circuit specialization results in faster executions and lower power consumption. 

Hardware specialization can offer a good trade-off between efficiency and flexibility when combined with reconfigurable computing. In this regard, \acp{fpga} are the most commonly used reconfigurable devices. They provide a fine granularity, which means outstanding levels of adaptivity but also high costs associated with the reconfiguration process. As an alternative, this paper targets reconfigurable hardware with coarser granularity, looking for a better balance between flexibility and reconfiguration overheads.

On this basis, \acp{cgra} are reconfigurable hardware accelerators that can compute domain-specific applications, as they can change coarse elements to perform a specific flow of arithmetic or control operations. ADRES~\cite{adres} and Morphosis~\cite{morphosys}, pioneering \acp{cgra}, laid the groundwork for exploring the potential of these architectures. \acp{cgra} are made of \acp{pe} forming an array, which contains \acp{fu} to perform operations and routing logic to interconnect with each other through direct paths or \acp{noc}. Configuration words are used to set the desired functionality of reconfigurable elements. Such configuration is generated by architecture-specific compilers, which are provided with \ac{dfg} extractors, schedulers, and mappers that place and route these graphs in the time and space dimensions. \acp{cgra} can be used for \ac{hpc}, competing against \acp{gpu} \cite{plasticine}, or low-power domains to compute workloads faster and more efficiently in embedded systems \cite{riptide}, targeting domains such as signal processing, security, and deep learning.

These reconfigurable fabrics are integrated with memories and control processors, providing flexible \acp{soc} featuring higher energy efficiency than traditional processing methods. This work proposes a \ac{cgra} microarchitecture design that supports offloading application sections with arithmetic operations and irregular loops. It uses elastic logic (i.e., valid and ready signals to make handshakes inside the \acp{pe} and the interconnection between them) to become latency tolerant, and it uses independent memory nodes to decouple address generation from \ac{cgra}. It is integrated into a \ac{soc} with a low-power \mbox{RISC-V} microcontroller following a streaming approach to transfer data between the main memory and the memory nodes inside the accelerator. The system is tested in post-synthesis simulations using TSMC 65 nm low-power technology and targeting a frequency of 250 MHz. It outperforms state-of-the-art works when processing data-oriented applications with regular access patterns. The accelerator obtains similar performance and energy efficiency results when computing applications with simple irregular loops, such as those in basic activation functions of \acp{nn}. The results are generated with signal processing, machine learning, and linear algebra benchmarks that cover cases where only one execution of the \ac{cgra} is needed (one-shot kernels) and more complex cases where the \ac{cgra} has to be reconfigured many times to complete the task (multi-shot kernels). The best results for one-shot kernels are a peak performance of 1.22 GOPs, an energy efficiency of 72.68 MOPs/mW, a speed-up of 17.63x, and a 9.05x of energy savings. The best results for multi-shot kernels are a peak performance of 1.17 GOPs, an energy efficiency of 115.96 MOPs/mW, a speed-up of 18.61x, and an 11.01x of energy savings.

The paper is structured as follows. Section~\ref{sec:background} presents the background and analyzes the state of the art. The main features of the microarchitecture are described in Section~\ref{sec:microarchitecture}, and the mapping strategies that can be used in it are presented in Section~\ref{sec:mapping-process}. The integration of the \ac{cgra} into a \mbox{RISC-V} \ac{soc} is presented in Section~\ref{sec:system-integration}. Section~\ref{sec:setup} outlines the experimental setup for the study, and the results and comparison with other works are shown in Section~\ref{sec:results}. Finally, the paper is concluded in Section~\ref{sec:conclusions}.

\section{Background}\label{sec:background}

This section provides a classification of \acp{cgra} focusing on how these fabrics exploit their spatial and temporal capabilities. It also exposes key features of some state-of-the-art works to differentiate them from the accelerator proposed in this paper.

\subsection{A taxonomy for \acp{cgra}}

Over the past two decades, different \ac{cgra} architectures tailored to specific application domains and design goals have been proposed. A taxonomy to classify the architectures by criteria such as their programming model, computation model, execution model, and microarchitecture was proposed in \cite{survey-taxonomy}. In \cite{survey-perf}, the architectures are classified by the target domain and the accelerators' performance. Differently, this background section focuses on how \acp{cgra} manage their processing capabilities (i.e., over time or space) and how the application mappings are scheduled, following the proposal in \cite{revel}.

Two scenarios from the application domain can be distinguished: \ac{hpc} and embedded devices. The two main differences are the size of the reconfigurable substrate and the amount of dedicated memory inside the accelerator. \acp{soc} in \ac{hpc} usually have less area and power consumption constraints. Therefore, architectures with hundreds of \acp{pe} and a considerable amount of memory can be implemented \cite{plasticine}. Low-power \acp{cgra} usually have smaller memories and tens of \acp{pe} \cite{ulp-srp}.

\acp{cgra} are spatial architectures that can achieve \ac{ilp}, by executing different instructions in different \acp{pe}, and \ac{dlp}, by executing the same operations to different data flows. Besides, \acp{cgra} can combine the space and the time to change the instruction a \ac{pe} executes in a given moment. Therefore, it can be distinguished between \acp{cgra} that only exploit their spatial capabilities \cite{dyser, dynaspam}, in which the configuration of each \ac{pe} remains unchanged until the execution of a kernel finishes, and \acp{cgra} that can receive different \ac{pe} configurations during execution \cite{adres, morphosys}. When using the former type, spatially-distributed \acp{cgra}, users or compilers only have to place and route the application \acp{dfg} into the architecture, while in the latter type, time-multiplexed \acp{cgra}, users or compilers have to recur to more complex techniques as the modulo scheduling~\cite{modulo-scheduling} to schedule instructions in each \ac{pe} and the flow of data inside the fabric. Therefore, spatially distributed \acp{cgra} are easier to use than time-multiplexed ones. From a hardware perspective, time-multiplexed \acp{cgra} need more logic to manage the change of instructions, usually using program counters and instruction memories.

Another key aspect of how a \ac{cgra} behaves is the type of scheduling it follows to propagate the data across the substrate. In statically-scheduled \acp{cgra} the data propagation is decided at compile time, balancing the paths to achieve data arriving at the right time to the corresponding \ac{fu}. In spatially-distributed \acp{cgra}, the substrate behaves as a systolic array \cite{stream-dataflow, tartan}, needing, in some cases, the insertion of data ``bubbles'' to delay a data path. In time-multiplexed ones, the modulo scheduling combines systolic management with the time multiplexing of instructions in the \acp{pe} \cite{adres, morphosys, hycube}. In contrast, dynamically-scheduled \acp{cgra} use dedicated logic to manage the data synchronization inside the fabric during execution time, relieving users or compilers of the task of routing compensation. This technique is also known as dataflow execution, referred to as static dataflow for spatially-distributed architectures and dynamic dataflow for time-multiplexed fabrics in the taxonomy of \cite{survey-taxonomy}. In the former, data transactions inside the \ac{cgra} follow a protocol to synchronize data consumption by the \acp{fu} \cite{dyser, plasticine}. In the latter type, the same technique is combined with tags to apply operations only to the proper data \cite{wavescalar, dmt-cgra}.

To achieve static dataflow computation in a \ac{cgra}, the use of elastic logic~\cite{synchro-inter-pipe, synt-synchro-elastic-arch} inside these architectures was proposed in \cite{elastic-cgras}. This approach relies on valid and ready signals to make the data exchange between producer and consumer effective, i.e., they become tolerant to latency. Some recent \acp{cgra} works that follow this philosophy are \cite{ue-cgra, riptide}. The main advantage of the elastic \acp{cgra} (static dataflow) is that they require less compilation and mapping effort compared with the other mentioned types, are more suitable in unpredictable latency scenarios, and reduce the routing logic needed to compute an application \cite{ss-vs-e}. Note that these advantages can lead to an overhead in area and energy consumption, as reported in \cite{exploration-elastic}.

\subsection{State-of-the-Art}

A selection of works related to the main microarchitectural concepts considered in this paper is presented in this subsection, including proposals that will later be used to compare qualitatively and quantitatively with the results obtained in this work.

In Softbrain~\cite{stream-dataflow}, authors propose the use of streaming engines with regular access patterns to simplify the offloading of kernels into the \ac{cgra}. In \mbox{dMT-CGRA}~\cite{dmt-cgra}, authors use multithreading techniques to exploit the accelerator parallelizing the tasks, targeting \ac{hpc} and comparing the work with \acp{gpu}. In REVEL~\cite{revel}, the use of \acp{cgra} as vector lanes of a \ac{vpu} is proposed, combining systolic computation with dataflow and using streaming engines to load and store data from the main memory. The authors include scratchpad memories within each lane and target \ac{hpc} using high frequency trying to compete with \ac{ooo} \acp{cpu}. In DSAGEN~\cite{dsagen}, a HW-SW codesign framework is provided to generate an architecture based on a set of applications that will be offloaded. In this work, authors also use streaming engines to communicate with memory. In CDA~\cite{cda}, authors apply unrolling of \ac{dfg} subgraphs to achieve more performance, targetting \ac{hpc} and comparing their results with other high-performance \acp{cgra} and \acp{gpu}. All these works have served as a reference to design the microarchitecture of the proposed \ac{cgra}.

Some state-of-the-art works already demonstrate the integration of a \ac{cgra} into an embedded \ac{soc}. These works are summarized here and will be used in Section \ref{sec:results} for comparison with the system proposed in this paper. IPA~\cite{ipa} is an ultra-low-power \ac{cgra} integrated with an OpenRISC processor that implements clock-gating mechanisms to improve energy efficiency. UE-CGRA~\cite{ue-cgra} uses an RV32IM \mbox{RISC-V} processor to control the \ac{cgra} accelerator and implements complex \ac{vlsi} mechanisms to change the clock frequency of the \ac{cgra} \acp{pe} to specialize in control-driven applications and reduce the energy consumption. RipTide~\cite{riptide} consists of an ultra-low-power \ac{cgra}, implemented with a \mbox{RISC-V} core with RV32EMC extensions. 

\section{\acs{cgra} Microarchitecture}\label{sec:microarchitecture}

This section proposes the microarchitecture of the elastic homogeneous \ac{cgra} that can compute data- and control-driven applications. It takes a baseline design as the starting point, which is first described to contextualize the original contributions of this work.

\subsection{Baseline design}

This work uses as the baseline \ac{cgra} microarchitecture the design presented by Capalija et al. in \cite{capalija}, provided as a high-performance \ac{fpga} overlay. The high-level structure of the \acp{pe} can be seen in Figure~\ref{fig:pe_structure}. Each \ac{pe} contains elastic logic to support latency-tolerant data interchanges using valid and ready signals. It has input and output ports in each cardinal position to communicate with the neighbors, two input ports and one output port in the \ac{fu}, the \ac{fu} that supports a fixed number of arithmetic operations, and versatile routing logic. The routing supports connections between the \ac{pe} input ports and the \ac{fu} inputs,  the \ac{pe} input ports and the \ac{pe} output ports, and the \ac{fu} output with \ac{pe} output ports. Furthermore, data can be accessed from multiple destinations. This design's \acp{pe} are connected in a mesh topology to the nearest neighbors (see Figure \ref{fig:cgra_structure}). The components of Figure~\ref{fig:pe_structure} contain the following logic in \cite{capalija}:

\begin{figure*}[t]
    \centering
    \begin{subfigure}{0.3\textwidth}
        \includegraphics[width=\textwidth,viewport=0 0 38mm 38mm]{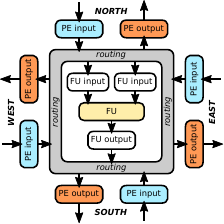}
        \caption{\ac{pe} structure.}
        \label{fig:pe_structure}
    \end{subfigure}
    \hspace{1cm}
    \begin{subfigure}{0.3\textwidth}
        \includegraphics[width=\textwidth, trim={1.75cm 1.75cm 1.75cm 1.75cm}, clip,viewport=0 0 133mm 136mm]{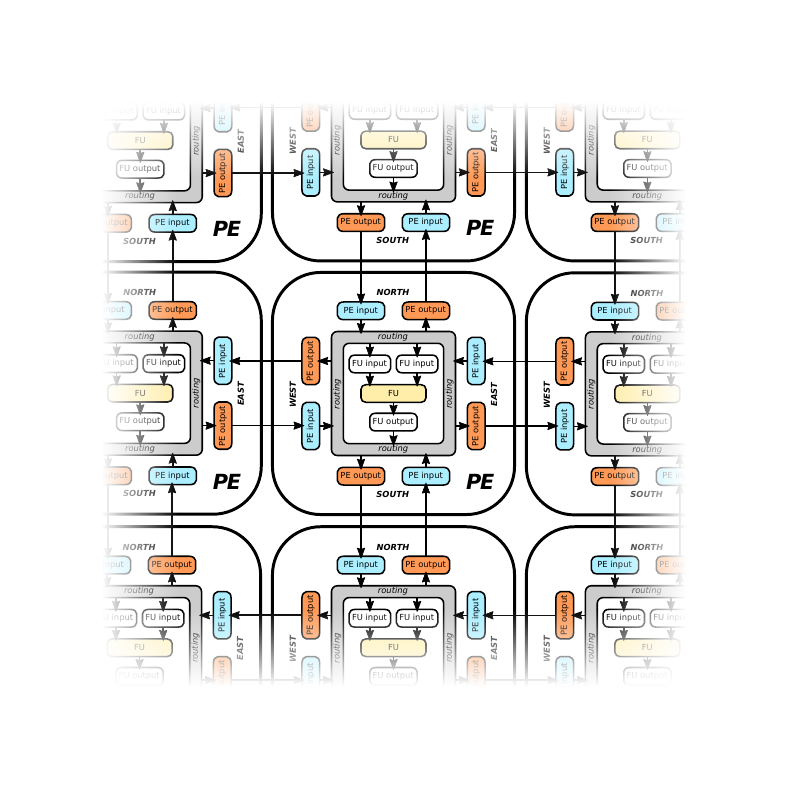}
        \caption{\ac{cgra} structure.}
        \label{fig:cgra_structure}
    \end{subfigure}
    \caption{Design structure.}
    \label{fig:design_structure}
\end{figure*}

\begin{itemize}
    \item \ac{pe} input ports: they have a \ac{fifo} memory with valid and ready signals connected to a \textit{Fork Sender}. This last element checks that all possible destinations (\ac{fu} inputs or \ac{pe} output ports) are ready to receive data.
    \item \ac{pe} output ports: contain a multiplexer to choose from which component the data come from, a \textit{Fork Receiver} to synchronize the data transfers in case there is more than one destination, and a simple valid/ready \ac{ff} that registers the data coming from the multiplexer and \textit{Fork Receiver}.
    \item \ac{fu} inputs: are similar to the \ac{pe} output ports, but changing the valid/ready \ac{fifo} for an \textit{Elastic Buffer} (behaves as a 2-slot \ac{fifo}). This component reduces the combinational paths and avoids combinational loops in all of his signals as it has two registers for data, two registers for the valid signal, and one register for the ready signal. 
    \item \ac{fu}: contains an elastic \textit{Join} component to commit the two operands with the same valid/ready signals and a \ac{alu} to perform integer or floating-point operations. The \ac{cgra} has heterogeneous distribution of operations, and each \ac{pe} has a limited set of operations. The integer design supports the following operations in their \acp{fu}: \textit{add}, \textit{sub}, \textit{abs}, \textit{mult}, and \textit{shift}. The floating-point design supports the following operations: \textit{add}, \textit{sub}, \textit{mult}, \textit{abs}, \textit{cmp}, \textit{div}, \textit{exp}, and \textit{log}.
    \item \ac{fu} output: are similar to the \ac{pe} input ports, but changing the \ac{fifo} for an \textit{Elastic Buffer}.
\end{itemize}

This design was extended by Zamacola et al. in \cite{zamacola} to add a constant as a possible operand of the \ac{fu}, logic to allow data reductions using a feedback loop inside the \ac{alu} of the \ac{fu}, support for bit-wise operations in the \acp{fu}, and other changes specific for the target of their work. Furthermore, this design follows a homogeneous distribution of operations inside the \ac{cgra}. The three former modifications introduced by Zamacola et al. will also be included in this work. Additionally, the homogeneous distribution will be used in this work for generality.

\subsection{Detected limitations}

The limitations of the baseline design are discussed in this subsection to justify the later proposals for modifications.

\subsubsection*{Design considerations}

The design uses simple valid/ready \ac{ff} in the \ac{pe} outputs to register data coming from a simple combinational path: the multiplexers that select if the data come from the \ac{fifo} of a \ac{pe} input or the \textit{Elastic Buffer} of the \ac{fu} output. If the timing constraints are not too tight, they could be removed. 

There are \textit{Fork Senders} in the \ac{pe} inputs and in the \ac{fu} output that are checking if all their enabled destinations are ready to receive data. In addition, there are \textit{Fork Receivers} in the \ac{fu} inputs and in the \ac{pe} outputs that also check if the other enabled destinations are available to coordinate data consumption. As they are doing the same task, one of these elements is redundant and thus can be removed.

\subsubsection*{Technology limitations}

When targeting the embedded domain, it is necessary to analyze the area, power, and energy aspects of the microarchitecture. The baseline \ac{cgra} targeted a \ac{fpga} implementation. They use \ac{fifo} memories as the input buffer of each \ac{pe}, which are inferred as block RAMs in the \ac{fpga} overlay, and one of their designs supported floating-point operations, which infer DSP blocks. In edge designs, implementing such memories and floating-point units would increase both the area occupancy and power consumption.

\subsubsection*{Functionality limitations}

The design only supports data-driven applications. The \ac{cgra} does not support offloading control-driven code sections, i.e., conditional statements and irregular loops. Embedded applications contain these types of cases, for instance, basic activation functions in \acp{nn} or simple image filters, which must be processed by the \ac{cpu}.

\subsection{Proposed microarchitecture}

This subsection presents the new microarchitecture based on \cite{capalija}. Design optimizations, embedded-domain adaptations, and major modifications to support irregular loops are integrated and employed in this new design. With respect to design optimizations, the following considerations are made:

\begin{itemize}
    \item The simple valid/ready \ac{ff} of each \ac{pe} output port is removed. After this modification, the maximum delay of the circuit is evaluated, ensuring that the target frequency constraint is still met (see synthesis results in Section \ref{sec:results}).
    \item The \textit{Fork Receivers} were removed because of redundancy, only the \textit{Fork Senders} are used in the design. These modules must be modified to assert the valid signal only if all the ready signals are set, using an AND gate that evaluates the valid and forked ready signals. 
\end{itemize}

Regarding the adaptation to the embedded domain, the following decisions were made:

\begin{itemize}
    \item \ac{fifo} memories of the \ac{pe} input ports are replaced by  \textit{Elastic Buffers} to lower the hardware resource utilization. This module registers the data and valid signals twice and the ready signal once. By buffering all the signals, the combinational paths are reduced, and the combinational loops are avoided.
    \item Regarding the data type and available operations, all the \acp{fu} of the design support the following integer data-oriented operations: \textit{add}, \textit{sub}, \textit{mult}, \textit{shift}, \textit{AND}, \textit{OR}, and \textit{XOR}.
\end{itemize}

To enable the execution of control-driven applications:

\begin{itemize}
    \item A comparator is included in the \ac{fu} to generate control signals inside the \ac{cgra}, which will be used to drive elements that manage conditional paths.
    \item \textit{Branch} and \textit{Merge}, elastic control components \cite{elastic-cgras}, are added to the \acp{fu}. The \textit{Branch} module has one data input and two possible data outputs that, depending on the value of the control signal data, the input is issued to one or the other. The \textit{Merge} module allows the confluence of the data coming from two different paths. In other words, the \textit{Merge} is reciprocal to the \textit{Branch}.
\end{itemize}

\begin{figure}[t]
    \centering
    \includegraphics[width=0.70\columnwidth,viewport=0 0 61mm 43mm]{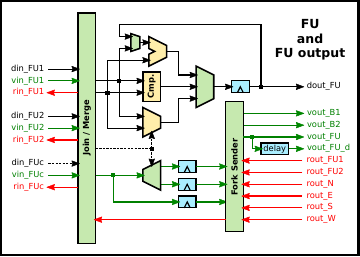}
    \caption{Microarchitecture of the \ac{fu} and its output.}
    \label{fig:fu_microarchitecture}
\end{figure}

The proposed microarchitecture is organized following the \ac{pe} structure of Figure \ref{fig:pe_structure}. The \ac{fu} and its output can be seen in Figure~\ref{fig:fu_microarchitecture}, which is divided into three parts: the \textit{Join/Merge} module, the datapath (in yellow), and the \ac{fu} output (registers with the \textit{Fork Sender}). The datapath (yellow components in Figure~\ref{fig:fu_microarchitecture}) has an \ac{alu}, a comparator, and a multiplexer. The \ac{alu} can compute \textit{add}, \textit{sub}, \textit{mult}, \textit{shift}, \textit{AND}, \textit{OR}, and \textit{XOR} operations. It can be noticed that one of its operands has a multiplexer to enable immediate feedback loops (data reductions). The comparator can compute the operations equal to zero and greater than zero needed to generate the control signal for the control-driven applications. Finally, the multiplexer allows the \textit{Merge} functionality and the computation of simple \textit{if/else} statements by using the multiplexer. The \textit{Join/Merge} has three inputs, two for the operands and one for the control signal, and it can operate in three modes:

\begin{itemize}
    \item \textit{Join} without control (two operand inputs active): this mode is intended for doing operations in the \ac{alu} or the comparator, where only the operands of the operation are needed.
    \item \textit{Join} with control (two operand inputs and the control input active): this mode is for using all the inputs of the \ac{fu}. It should be selected to perform a \textit{Branch} or to use the multiplexer as a \textit{if/else} module. The \textit{Branch} uses the operand inputs in the \ac{alu} or the comparator, and the control input to generate the \textit{Branch} valid signals (\textit{vout\_B1} and \textit{vout\_B1}), which will be the output valid signal for different cardinal positions. To perform an \textit{if/else} statement, the two operands are the inputs of the datapath multiplexer, being the control signal the selector of this module.
    \item \textit{Merge} (two operand inputs active): in this mode, there are two possible operand inputs, but only one of them will commit data at the same time (we want to confluence two paths in one). The control signal does not come from outside the \ac{fu}, it is generated internally to drive the datapath multiplexer.
\end{itemize}

The three outputs of the datapath are multiplexed to choose one of them, and then, the output is registered to enable the immediate feedback loop and to reduce the combinational paths as much as possible. The \ac{fu} output is in charge of the management of the valid and ready signals. Three valid signals (registered to balance with the data path) are the input of the \textit{Fork Sender}: the unprocessed valid coming from the \textit{Join/Merge} and two valid signals that are demultiplexed using the control signal to implement the \textit{Branch} functionality. The forked ready signal goes straight to the \textit{Join/Merge}. After the \textit{Fork Sender}, four valid signals are available: the unprocessed valid (\textit{vout\_FU}), a delayed unprocessed valid to commit data reductions or terminate loops (\textit{vout\_FU\_d}), and the two valid signals of the \textit{Branch} (\textit{vout\_B1} and \textit{vout\_B2}). The ready signals of the possible destinations of its output are the non-immediate feedback loops (\textit{rout\_FU1} and \textit{rout\_FU2}) and the cardinal \ac{pe} outputs (\textit{rout\_\{N,E,S,W\}}).

\begin{figure}[t]
    \centering
    \includegraphics[width=0.90\columnwidth,viewport=0 0 79mm 47mm]{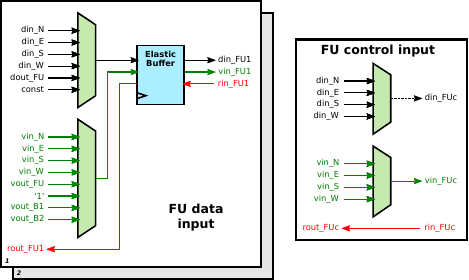}
    \caption{Microarchitecture of the \ac{fu} data and control inputs.}
    \label{fig:fu_io_microarchitecture}
\end{figure}

There are three inputs to the \ac{fu}, two for data and one for the control signal. Its microarchitecture is shown in Figure~\ref{fig:fu_io_microarchitecture}. Both \ac{fu} data inputs have a data and valid multiplexer to select the origin of the operand. Finally, an \textit{Elastic Buffer} avoids combinational loops with the ready signal in case of a non-immediate feedback loop (data coming from \textit{dout\_FU}). In contrast, the \ac{fu} control input only supports data coming from the \ac{pe} inputs, and it does not need an \textit{Elastic Buffer} because non-immediate feedback loops are not allowed for control. The control signal always comes from outside. Thus, the ready signal cannot produce a combinational loop.

\begin{figure}[b]
    \centering
    \includegraphics[width=0.95\columnwidth,viewport=0 0 86mm 46mm]{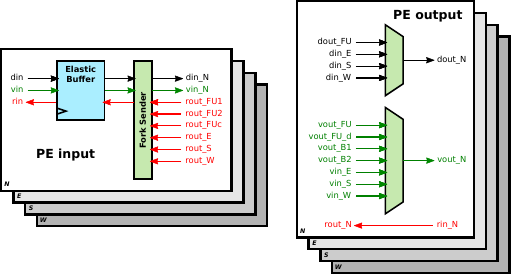}
    \caption{Microarchitecture of the \ac{pe} inputs and outputs.}
    \label{fig:pe_io_microarchitecture}
\end{figure}

The four \ac{pe} input ports contain an \textit{Elastic Buffer} connected to a \textit{Fork Sender}. They receive ready signals from their possible destinations: the three inputs of the \ac{fu} (2 for data and 1 for control) and the three direct outputs in the \ac{pe}. The four \ac{pe} output ports just contain data and valid multiplexers to select the origin of the data. Their microarchitecture can be seen in Figure~\ref{fig:pe_io_microarchitecture}.

The presented components are reconfigurable to achieve a given functionality in the \ac{cgra}, which is set through a \ac{pe} configuration word. The reconfigurable parameters of the \ac{fu} are the operation of the \ac{alu}, the select signal of the immediate feedback loop multiplexer of the \ac{alu}, the operation of the comparator, the \textit{Join/Merge} mode, the select signal of the final datapath multiplexer, the initial value of the data register, the initial values of the three valid registers, a mask for the \textit{Fork Sender}, and the value for the delay of the non-processed valid signal. The initial values of the registers inside the \ac{fu} are employed to start a flow inside the \ac{cgra}. The reconfigurable parameters of the \ac{fu} inputs are the six select signals of their multiplexers and the constant value. The reconfigurable values of the \ac{pe} input ports are the masks of the \textit{Fork Senders}. Finally, the reconfigurable parameters of the \ac{pe} output ports are the select signals of their multiplexers. 

All these configuration fields sum up 144 bits per \ac{pe}, and all the active \acp{pe} of a kernel must be configured before the execution of the \ac{cgra}. Each word can set a \ac{pe} to behave as an \ac{alu}, \textit{Branch}, \textit{Merge}, or \textit{if/else} cell, also allowing to specify initial values of the registers inside the \ac{fu} so that counters or accumulators can be initialized.

\section{Mapping process}\label{sec:mapping-process}

In this section, the mapping process is introduced. Before mapping an application section, a \ac{dfg} of the section of code that will be offloaded is extracted, and then, the operations are placed according to the routing availability, following the architectural and mapping considerations presented next.

\subsection{Architectural considerations}

\begin{figure}[H]
    \centering
    \includegraphics[width=0.75\columnwidth,viewport=0 0 68mm 53mm]{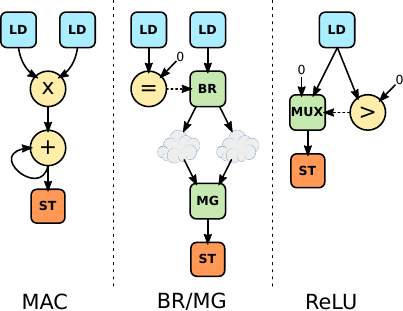}
    \caption{\ac{dfg} examples.}
    \label{fig:dfgs}
\end{figure}

As mentioned in Section~\ref{sec:microarchitecture}, the \ac{cgra} can compute data- and control-driven applications. When mapping applications on it, the following differences in the use of the \acp{fu} are found:

\begin{itemize}
    \item Data-driven applications use the \ac{alu} of the \acp{fu}, with the optional immediate feedback loop, and the \textit{Join/Merge} configured with the ``\textit{Join} without control'' mode. The two possible valid signals are the unprocessed \textit{vout\_FU} and the delayed unprocessed \textit{vout\_FU\_d}. An example of these types of applications can be seen in the \ac{dfg} on the left of Figure~\ref{fig:dfgs} (MAC). 
    \item Control-driven applications also can use the \acp{alu}, although they need one or more comparators and one or more of the available control-flow modules: \textit{Branch}, \textit{Merge}, or the multiplexer. To set the \textit{Branch} functionality, the \textit{Join/Merge} module has to be configured in ``\textit{Join} with control'' mode to use the control signal to drive the demultiplexer of the valid signals, and the \textit{vout\_B1} and \textit{vout\_B2} signals have to be routed to two different flows of the \ac{dfg}. To use the \ac{fu} as a \textit{Merge}, the \textit{Join/Merge} module has to be set with the ``\textit{Merge}'' mode, which is in charge of generating the control signal to drive the datapath multiplexer. On the center \ac{dfg} of Figure~\ref{fig:dfgs} (BR/MG), an example of a \textit{Branch} and a \textit{Merge} can be seen. The last control-flow method is to use the multiplexer in a ``traditional'' way. The \textit{Join/Merge} module should be working in ``\textit{Join} with control'' mode to use the control input signal as the selector of the datapath multiplexer, allowing the mapping of \acp{dfg} as the one on the right of Figure~\ref{fig:dfgs} (ReLU).
\end{itemize}

It has to be noticed that the comparisons should be made in isolated \acp{pe}, not being possible to drive a control-flow module (\textit{Branch}, \textit{Merge}, or multiplexer) in the same \ac{pe}.

\subsection{Mapping considerations}

The input data operands and the output results of the computation in the \ac{cgra} should be placed in the borders of the fabric. To maximize the space to host the kernels and even replicate the \acp{dfg} to apply unrolling, it is established that the inputs will be positioned in all the \acp{pe} of a \ac{cgra} border, and the outputs will be in the opposite border. In particular, the inputs are placed on the north border and outputs on the south border of this \ac{cgra}. The east and west borders, except in the first and last row, are used as complementary south-to-north paths because it has been found that these are the more congested routes. Thus, the \ac{cgra} now has \textit{n} south-to-north paths plus two more. Besides this, the dimensions of the \ac{cgra} often restrict the ability to compute complete applications at once. Three different ways to map applications, referred to as mapping strategies, are distinguished:

\begin{enumerate}
    \item The kernel is small enough to fit in the architecture and big enough to use most of the resources in the \ac{cgra}.
    \item The kernel is so small that it can be unrolled several times within the \ac{cgra}, incrementing the throughput.
    \item The kernel is too large and does not fit into the \ac{cgra}. In this case, algorithms must be split into small chunks, which can be offloaded using the two former mapping strategies. This way, the accelerator is reconfigured multiple times to execute different kernels or data for computing the whole application.
\end{enumerate}

The first two mapping strategies are considered one-shot kernels because they only need one \ac{cgra} configuration and execution to finish the application, and the third mapping strategy is referred to as multi-shot kernels for relaunching and reconfiguring the substrate several times.

\section{System Integration}\label{sec:system-integration}

The proposed \ac{cgra} has been integrated with a RISC-V microcontroller based on \ac{xheep}~\cite{xheep}, forming a \ac{soc}, including memory resources and peripherals to allow real-world applications. In this context, \ac{cgra} will serve as a general-purpose accelerator, enabling offloading computing-intensive code sections from the processor. This section describes the selected \ac{soc} and all the logic to control and feed the accelerator.

\subsection{\ac{xheep}}

\ac{xheep}~\cite{xheep} is an open-source platform designed to support the integration of ultra-low-power edge accelerators. It provides customization options to match specific application requirements by exploring various core types, bus topologies, and memory size and addressing modes. It also enables a fine-grained configuration of memory banks to be aligned with the constraints of the integrated accelerators. The platform prioritizes energy efficiency by implementing low-power strategies, such as clock-gating and power-gating, and integrating these with connected accelerators through dedicated power control interfaces. The user can select the RISC-V among three OpenHW Group \acp{ip}: CV32E20, CV32E40P, and CV32E40X. Such a selection is based on the required trade-off between power and performance. In particular, the CV32E20 core is optimized for control-oriented tasks, while the CV32E40P core is optimized for processing-oriented tasks \cite{schiavone2017slow}. The CV32E40X core offers power consumption and performance similar to the CV32E40P core without featuring floating-point RVF and custom Xpulp \ac{isa} extensions and exposing an interface (eXtendable-interface) to extend the \ac{isa} easily. Regarding the bus topologies, either a one-at-a-time topology, where only one master can access the bus at a time, or a fully connected topology, where multiple masters can access multiple slaves in parallel, can be configured to trade area with bandwidth. A different number of memory banks can be selected to trade off area, power, and storage capacity for memory configuration. Each bank offers a retention state to reduce leakage power when the bank is not accessed for some time, but the data needs to be preserved. Finally, the peripherals include a platform-level interrupt controller (PLIC), a timer, and general-purpose I/O peripherals such as a GPIO, an I2C, I2S, an SPI, a DMA, a UART, etc.

% Justification of choice
As most of the features of the \ac{xheep} platform are oriented toward reducing energy consumption, relying on it will ease the design and implementation of clock gating and power gating strategies also for the \ac{cgra}, which will be presented in Section~\ref{sec:energy-saving-mechanisms}. Furthermore, it is an open hardware project that allows the customization of its internal components. For this work, the authors have contributed to the \ac{xheep} project by adding an optional interleaved bus. This feature allows simultaneous access to a memory range as the least significant part of the memory addresses determines the bank where data is stored, and the number of interleaved banks can be selected between 2, 4, and 8. The interleaved bus increases the memory bandwidth between main memory and hardware accelerators with multiple memory ports, as every master node works in parallel on a different memory bank, reducing the probability of conflicts when the application allows~\cite{interleavedmemory}. As described in the following subsection, independent memory nodes in the \ac{cgra} are needed to exploit this connection.

\subsection{Platform and accelerator}\label{sec:accelerator}

The \ac{cgra} proposed in Section~\ref{sec:microarchitecture} is intended for offloading intensive sections of code on it from the processor. The \ac{cpu} controls the \ac{cgra} and shares the memory banks of the \ac{soc} with it. Figure \ref{fig:block_diagram_cgra-x-heep} shows the block diagram of the \ac{cgra} accelerator. The control unit receives requests from the \ac{cpu} through a memory-mapped interface and orchestrates the entire accelerator's behavior with the received commands. \acp{imn} can be seen in blue and \acp{omn} in orange, all receiving data from the control unit and interconnecting with the system bus crossbar of \ac{xheep} to load and store data, respectively. The core of the design is the \ac{cgra}, which receives data and configuration from the \acp{imn} and delivers the processed data to the \acp{omn}.

\subsubsection*{Input/Output Memory Nodes}

\begin{figure}[t]
    \centering
    \includegraphics[width=0.9\columnwidth,viewport=0 0 119mm 141mm]{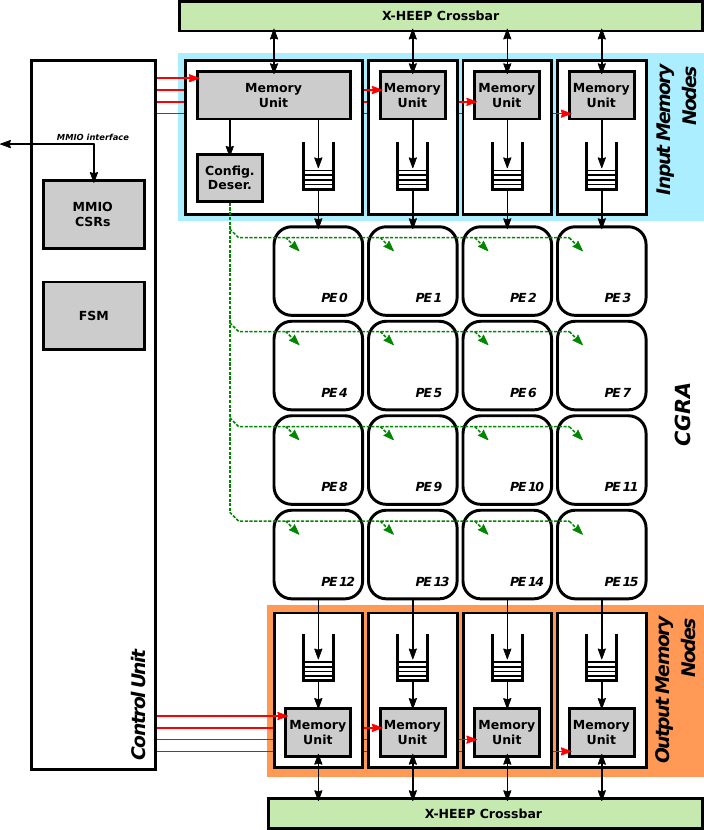}
    \caption{Block diagram of the \acs{cgra} accelerator.}
    \label{fig:block_diagram_cgra-x-heep}
\end{figure}

Multiple master input/output memory nodes have been included in the CGRA IP to take advantage of the interleaved accesses, allowing multiple data transfers per cycle that depend on the number of interleaved banks. \acp{imn} request data streams or scalars to memory and inject them in the \ac{cgra} to process a specific kernel. The first \ac{imn}, see Figure~\ref{fig:block_diagram_cgra-x-heep}, also fetches the kernel configuration words from memory. In this work, a 6-bit identification number is added to each \ac{pe} configuration word to know to which \ac{pe} the configuration is for. This makes it possible to have variable-size kernel configurations. Then, a deserializer joins the five 32-bit words to form the 152-bit configuration word that needs each \ac{pe}. \acp{omn} store the processed data streams or scalars in memory. This streaming approach was proposed in Softbrain~\cite{stream-dataflow} to combine the dataflow computation of a \ac{cgra} and stream memory engines. As can be seen in Figure~\ref{fig:block_diagram_cgra-x-heep}, the nodes contain memory units that generate the new stream addresses using three parameters written by the \ac{cpu}: the initial address, the size of the stream, and the associated stride of the stream. Thus, the \acp{pe} are leveraged from this duty to process only operations of the kernel \ac{dfg}. They also contain \ac{fifo} memories between the memory units and \ac{cgra} to dampen data transfers in case of stalling, which can happen if there are more active memory nodes than the number of interleaved banks.

\subsubsection*{CGRA memory Infrastructure}

Another aspect related to the memory infrastructure to consider is the presence of dedicated data memories inside the accelerator. Softbrain~\cite{stream-dataflow} uses a private memory to avoid redundant data accesses, supporting data streamings to both main and private memory. In contrast, Riptide~\cite{riptide} does not implement it and takes advantage of its latency-tolerant nature to directly request data loads or stores to the main memory. This design decision determines the maximum quantity of data that can be processed at once, and it has area, performance, and power implications. The \ac{cgra} accelerator proposed in this work does not use a private memory for the following reasons:

\begin{enumerate}
    \item The size of a dedicated memory would limit the amount of data that can be processed in one \ac{cgra} execution, while the memory nodes can request vectors of a larger size.
    \item The design is area-constrained, and a dedicated memory in the accelerator would increase the costs.
    \item There are situations where performance and power consumption can be better when dedicated memories are used, typically when redundant memory accesses exist. When data can be reused from the dedicated memory, performance is higher and power consumption is lower because the bandwidth requested to the main memory is lower than when no private memories are used, easily saturating the memory infrastructure. Despite this improvement when this case happens, the two former points are considered more important.
\end{enumerate}

\subsubsection*{CGRA Access from the Processor}

The accelerator control unit is in charge of communication with the \ac{cpu}. This \ac{ip} has memory-mapped registers to configure \acp{imn} and \acp{omn}. Configuration information, data information, and start and control commands should be provided. Once a kernel execution is performed, an interrupt signal is sent to the \ac{cpu}.

\subsubsection*{Conputation Model}

The computation model proposed for the integrated platform combines memory nodes to stream data into the elastic \ac{cgra}. For this reason, it receives the name of \ac{strela} \ac{cgra}. The first time the \ac{cgra} is configured to perform a given kernel, or it is needed to change it, the address and size of the kernel configuration should be provided. In that case, the architecture must spend some clock cycles, which depend on the number of \acp{pe} that a kernel uses, fetching this reconfiguration stream. Then, the data streams should be configured to load and store data into the \ac{cgra}. Therefore, the preamble of every kernel execution consists of write operations to memory-mapped registers. Once configured, the hardware can execute the kernel. The loading, execution, and storing of data are done at the same time because the memory nodes and the \ac{cgra} core are independent and tolerant to latency. The achieved performance will depend on the \ac{ii} that a specific kernel produces in the \ac{cgra} and the number of interleaved banks (bandwidth available). The ideal case is an \ac{ii} of one with enough bandwidth to make the required loads and stores at the same time. All these features provide the accelerator with elements that make it ready to use in real projects, avoiding, in most cases, bottlenecks in the memory infrastructure, the most penalizing part of hardware acceleration.

\subsection{Energy-saving mechanisms}\label{sec:energy-saving-mechanisms}

Since it targets the embedded domain, the \ac{cgra} area must be as small as possible to reduce manufacturing costs and power consumption. Clock and power gates are hardcoded at different hierarchy levels to adapt the \ac{cgra} to the low-power domain. These features are complementary to the automatic clock-gating set at the synthesis configuration in the \ac{eda} tool. The full accelerator can be clock- or power-gated at the first level. The system software handles these procedures to switch off the supply voltage or the clock of the accelerator when unused, and they are controlled by the power manager of \ac{xheep}. The clock is only enabled in the \ac{pe} matrix at the second level when the system executes a kernel. This procedure is controlled by hardware inside the accelerator and saves energy by only maintaining enabled the \acp{csr} of the \ac{cgra} at idle status. The lower stages of the clock-gating hierarchy act at \ac{pe} and \textit{Elastic Buffer} level. If a \ac{pe} is provided with a configuration word, the clock is automatically enabled for this element. This clock-gating relies on the system software because it entirely depends on the kernel configuration data. Finally, the \textit{Elastic Buffers} inside each \ac{pe} are clock-gated within a specific field in the configuration word to have active only the used border crossings and used operands. To do so, six bits are added to the 152-bit configuration word, so that each \ac{pe} needs 158 bits of configuration (144 bits for reconfigurable elements, 6 bits for the ID, and 6 bits for the clock gating). Therefore, these switches are also managed by software.

\section{Experimental setup}\label{sec:setup}

In this section, the \ac{soc} configurations for testing the design are chosen, and evaluation tools are presented. Then, some representative benchmarks are selected to exploit the accelerator in the different scenarios presented in Section~\ref{sec:mapping-process}, giving some examples of manual mappings to understand the compilation and mapping process better.

\subsection{System implementation}\label{sec:system-implementation}

Among the existing possibilities of \ac{xheep}, the \textit{CV32E40P} core~\cite{cv32e40p}, an energy-efficient, 32-bit, in-order RISC-V core with a 4-stage pipeline that implements the RV32IMC \ac{isa} has been selected as the main processor in this work. To exploit the potential of the design properly, the multi-master-at-a-time access system bus configuration (\textit{NtoM}) has also been selected. Eight banks of 32 KB (the first four with continuous addressing and the last four with interleaved addressing) are used for memory. The interleaved section of the memory subsystem gives up to 128 bits/cycle of bandwidth when four or more masters request data simultaneously. The \ac{strela} accelerator is configured to have a 4x4 \ac{pe} array with a 32-bit datapath.

The final design targets a maximum frequency of 250 MHz in the worst-case scenario (slow-slow corner, 1.08~V, 125~ºC). It is synthesized with Synopsys Design Compiler 20.9 using TSMC 65 nm Low-Power technology libraries (only low-threshold voltage cells), configured with multi-power domains to enable the previously mentioned power-gating strategy. The RTL is simulated with Mentor Questasim 20.04, and the post-synthesis consumption metrics are generated with Synopsis PrimePower 22.12.

\subsection{Benchmarks}\label{sec:benchmarks}

Applications to test the architecture have been selected to demonstrate the ability of \ac{cgra} to map both control and data-driven applications and the capacity to map kernels that fit entirely and partially into the reconfigurable fabric, as described in Section~\ref{sec:mapping-process}. Furthermore, the selected benchmarks must represent the embedded domain, for instance, signal and image processing, deep learning, and linear algebra. This demonstrates the generality of the proposal and that the execution of target applications in the \ac{cgra} improves the performance and energy efficiency compared to the base \ac{xheep} without the custom design.

The one-shot data-driven selected application is \textit{fft}, which implements a 2-radix butterfly diagram. The one-shot control-driven selected applications are the \textit{ReLU} activation function, the \textit{dither} image filter used in \cite{ue-cgra}, and the \textit{find2min}, which finds two minimums and its indexes from a list (used to find valleys in heart pulse signals). These benchmarks are representative of the different behavior of \ac{cgra} when running an iteration with both data and control-driven requirements but also of exploring the impact of the two first mapping strategies, since \textit{ReLU} is small enough to have an unrolling of 3 and \textit{dither} an unrolling of 2.

For the multi-shot kernels, only data-driven applications are considered. Control-driven applications typically have a \ac{ii} greater than one because they rely on many control signals that require additional time to propagate, meaning they take considerable time to execute. On the other hand, data-driven applications have larger \acp{dfg} and rely less on control signals, making them more suitable for the multi-shot kernels study. The \textit{conv2d}, 2-dimensional convolution present in \acp{cnn}, is selected because it has a fixed amount of iterations, 3 in total (one for each row of the $3\times3$ filter). In contrast, \textit{mm}, dense matrix multiplication benchmark, is selected because it needs a variable number of executions that depend on the matrix dimensions. Finally, we use Polybench \cite{polybench} benchmarks to test the architecture with linear algebra applications: \textit{gemm}, \textit{gemver}, \textit{gesummv}, \textit{2mm} and \textit{3mm}.

\begin{figure*}[t]
    \centering
    \begin{subfigure}{0.3\textwidth}
        \includegraphics[width=\textwidth,viewport=0 0 73mm 75mm]{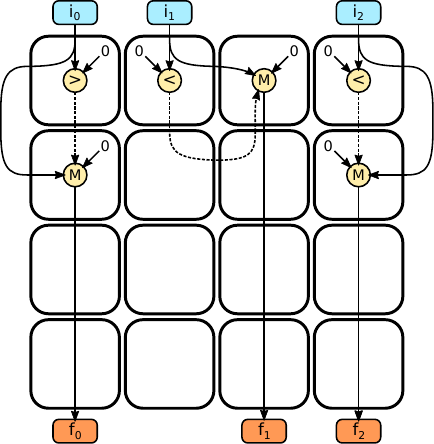}
        \caption{ReLU mapping (unrolling of 3).}
        \label{fig:mapping-relu}
    \end{subfigure}
    \hfill
    \begin{subfigure}{0.3\textwidth}
        \includegraphics[width=\textwidth,viewport=0 0 73mm 75mm]{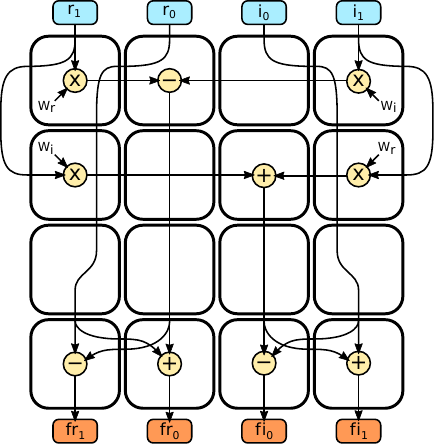}
        \caption{FFT mapping (full).}
        \label{fig:mapping-fft}
    \end{subfigure}
    \hfill
    \begin{subfigure}{0.3\textwidth}
        \includegraphics[width=\textwidth,viewport=0 0 73mm 75mm]{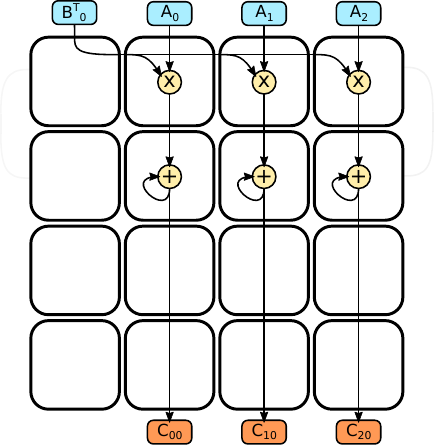}
        \caption{MM mapping (partial).}
        \label{fig:mapping-mm}
    \end{subfigure}
            
    \caption{Manual \ac{cgra} kernel mappings.}
    \label{fig:cgra-mappings}
\end{figure*}

All these benchmarks have been mapped manually into the \ac{cgra}, following the strategies mentioned in \ref{sec:mapping-process}, because this paper focuses on the hardware side of the system. In Fig. \ref{fig:mapping-relu}, unrolling has been applied to replicate the \ac{dfg} of \textit{ReLU} (shown in Fig. \ref{fig:dfgs}) only three times due to congestion. The maximum unrolling is 4 when the routing allows it. The mapping of the \textit{fft} kernel can be seen in Fig. \ref{fig:mapping-fft}. With this benchmark, the \ac{cgra} accelerator is fully utilized as all memory nodes and all \acp{pe} are used, the routes going from north to south are congested, and it has many \acp{fu} doing operations. Finally, in Fig. \ref{fig:mapping-mm}, a partial kernel that computes three dot products at a time is shown (\ac{dfg} of the MAC operations present in this computation is shown in Fig. \ref{fig:dfgs}). This mapping is part of the \textit{mm} benchmark, which needs to be deployed the number of columns times the number of rows divided by 3 (as it has an unrolling of 3). It can be seen that every time an intermediate execution is finished, new vector addresses should be configured in the \ac{cgra} to compute the whole matrix multiplication.

\section{Experimental Results}\label{sec:results}

In this section, the experimental results of the \ac{cgra} accelerator integrated into the \ac{xheep} \ac{soc} are presented. First, synthesis results are shown for the \ac{soc} configuration mentioned in \ref{sec:system-implementation}. Then, the architecture is evaluated with the two benchmark types presented in \ref{sec:benchmarks}. These results are discussed to showcase the characteristics of this implementation. Finally, this work is compared with the most relevant state-of-the-art \ac{cgra} works to show qualitative and quantitative differences.

\subsection{Synthesis results}

\begin{figure}[b]
    \centering
    \includegraphics[width=0.99\columnwidth,viewport=0 0 249mm 89mm]{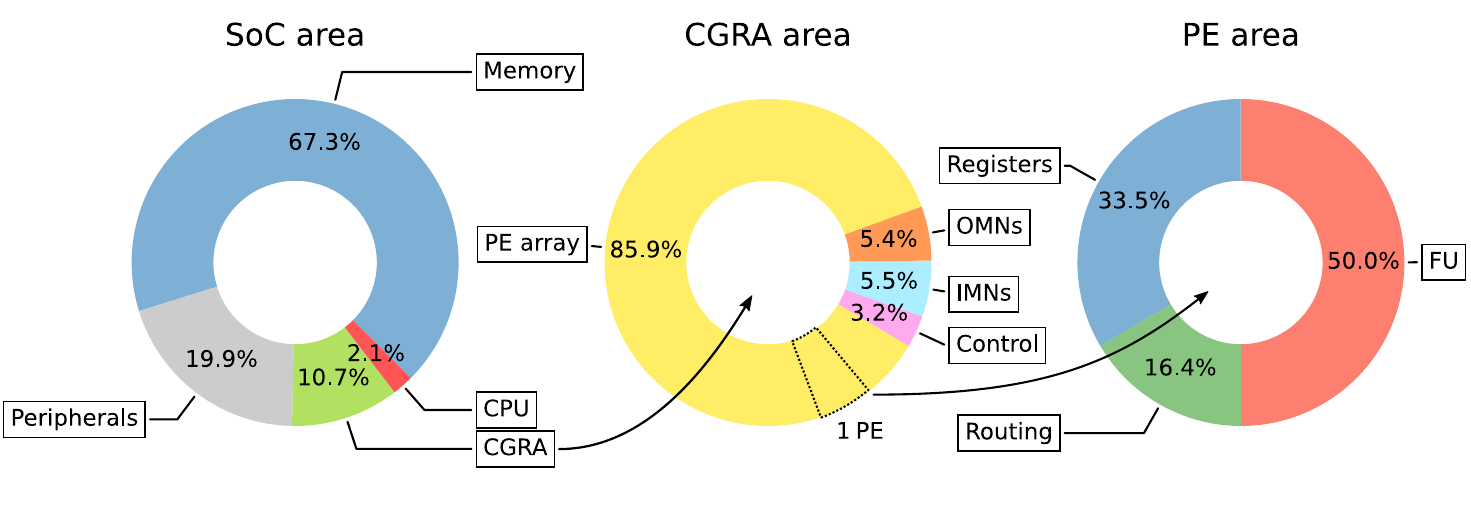}
    \caption{Synthesis area percentage results.}
    \label{fig:area}
\end{figure}

The \ac{cgra} area is 13,936 $\mu m^2$ for each \ac{pe}, 253,442 $\mu m^2$ for the entire \ac{cgra} accelerator, and 2,38 $mm^2$ for the \ac{soc}. Fig. \ref{fig:area} shows the area percentages for the different components. The elastic logic of this design introduces some overhead to the area utilization, as reported in previous state-of-the-art works dealing with the adaptation of inelastic architectures ~\cite{ue-cgra, exploration-elastic}. Despite this overhead, the \acp{fu} are the most area-consuming, and they do not belong to the elastic part of the architecture. They have a 1-cycle datapath (\ac{alu}, comparator, and multiplexer) that supports arithmetic and control operations. This \ac{pe} reported area is similar to other work using the same node size ~\cite{cma}. The area of the whole accelerator system is mainly the \ac{pe} matrix area, with the control logic, \acp{imn}, and \acp{omn} consuming 14.1 \% of the accelerator area. On the pie chart on the right, it can be seen the \ac{soc} area utilization. The 256 KB memory is the most area-consuming part, with 67.3 \% of the total. The \ac{cgra} takes about five times the area the single \ac{cpu} selected in this work uses.

\subsection{Performance and power consumption results}

Using the benchmarks presented in \ref{sec:benchmarks}, performance and consumption metrics are reported for both the called one-shot and multi-shot kernels. The \ac{soc} is the same in all the tests, and the results are generated in post-synthesis simulations. When reporting \ac{cpu} metrics, the \ac{cgra} accelerator is clock- and power-gated. When reporting the \ac{cgra} metrics instead, less power optimizations are enabled since the \ac{cpu} acts as the control core. The compiler is always set with the ``-O3'' optimization flag. Performance results depend on the number of operations considered, which can differ when evaluating different architectures. In this paper, only arithmetic operations are considered to provide architecture-agnostic metrics. For instance, in \textit{fft} (see Fig. \ref{fig:mapping-fft}), ten arithmetic operations are performed every four inputs. The formula ~$2n^3-n^2$ provides the total operations for the naive matrix multiplication algorithm. In the case of control-driven applications, counting operations can be complex as multiple paths are possible (branch functionality), although only one is effective at each time. For simplicity, all the enabled \acp{fu} are counted as operations in these benchmarks. As the operation count may not be trivial in control-driven applications, another architecture-agnostic performance metric of output per cycle is included. As discussed in \cite{riptide}, it is difficult to find other state-of-the-art works that provide results considering this. For this reason, the performance and energy efficiency comparison made in the following subsection only includes three works.

\begin{table*}[t]
    \centering
    \noindent
    \caption{One-shot kernel results}
    \begin{tabular}{r | N N N N }
        \toprule
        \textbf{Kernel} & \textbf{\textit{fft}} & \textbf{\textit{relu}} & \textbf{\textit{dither}} & \textbf{\textit{find2min}} \\
        \cmidrule(lr){1-5}
        \textbf{Configuration cycles} & 84 & 74 & 74 & 84 \\
        \textbf{Execution cycles} & 523 & 697 & 4,617 & 7,175 \\
        \textbf{Number of operations} & 2,560 & 2,048 & 5,120 & 9,216 \\
        \textbf{Outputs/cycle} & 1.95 & 1.47 & 2.22$\times\num{e-1}$ & 5.57$\times\num{e-4}$ \\
        \cmidrule(lr){1-5}
        \textbf{Performance (MOPs)} & 1,223.71 & 734.58 & 277.24 & 321.11 \\
        \textbf{CGRA consumption (mW)} & 16,84 & 11.51 & 9.01 & 9.64 \\
        \textbf{Energy efficiency (MOPs/mW)} & 72.68 & 63.80 & 30.76 & 33.31 \\
        \cmidrule(lr){1-5}
        \textbf{CPU cycles [-O3]} & 9,218 & 10,759 & 14,342 & 14,381 \\
        \textbf{CPU consumption (mW)} & 4.04 & 3.44 & 3.54 & 3.37 \\
        \cmidrule(lr){1-5}
        \textbf{Speed-up} & 17.63x & 15.44x & 3.11x & 2.00x \\
        \textbf{Energy savings (CPU vs. CGRA)} & 4.23x & 4.62x & 1.22x & 0.70x \\
        \cmidrule(lr){1-5}
        \textbf{SoC CGRA consumption (mW)} & 53.84 & 45.34 & 28.84 & 28.84 \\
        \textbf{SoC CPU consumption (mW)} & 27.59 & 26.59 & 26.09 & 26.59 \\
        \textbf{Energy savings (SoCs)} & 9.03x & 9.05x & 2.81x & 1.85x \\
        \bottomrule
    \end{tabular}
    \label{tab:one-shot_kernels}
\end{table*}

Table \ref{tab:one-shot_kernels} shows results for the one-shot kernels. The total amount of input data is set to 1024, having for \textit{fft} input vectors of 256 data each (4 input ports active), 341 for \textit{relu} (3 input ports active as it has an unrolling of 3), 512 for \textit{dither} (2 input ports active as it has an unrolling of 2), and 1024 for \textit{find2min} (just one input port active). To offload kernels in the \ac{cgra}, all their needed \acp{pe} should have their configuration word loaded in advance. This is only needed the first time a kernel is run or whenever a new configuration has to be set. The configuration cycles are different for each kernel as they depend on the number of \acp{pe} used in a particular kernel. Furthermore, some control overhead appears when the vector addresses and sizes are loaded in the memory-mapped registers and during synchronization with the processor when the accelerator is done. These preamble cycles are not used in the performance metrics of the one-shot kernels (they are used in the multi-shot kernels) to show the better result that tends to occur when increasing the input data size when the control and reconfiguration time can be neglected. Performance results show that the data-driven application (\textit{fft}) obtains better values than control-driven applications (\textit{relu}, \textit{dither}, and \textit{find2min}). \textit{fft} does not present a \ac{ii} greater than one because the \ac{dfg} does not have any feedback data dependency in the graph, and the bottleneck is in the bus subsystem of the \ac{soc}. Having four interleaved banks and 128 bits/cycle of bus bandwidth, the eight memory nodes of the \ac{cgra} requesting simultaneously 256 bits/cycle are accomplished in more than one cycle (ideally two clock cycles), thus the \ac{cgra} achieves 1.95 outputs per cycle and not 4. In contrast, some control-driven applications have feedback data dependencies in the graph that produce an \ac{ii} on the \ac{cgra} data inputs. These conditions occur in \textit{dither} and \textit{find2min}, with an \ac{ii} of 4 and 6, respectively. In \textit{relu}, the performance is better because there is no data dependency producing feedback in the \ac{dfg}, and the mapping is unrolled, causing the bottleneck on the bus again. Regarding power consumption, \acp{pe} that perform arithmetic operations consume more than the control or routing ones. Also, the number of used memory nodes increases the metric. This is why there is a difference in reflected consumption between data-driven and control-driven benchmarks. Finally, the energy savings are reported comparing the \ac{cgra} vs. the \ac{cpu}, and at \ac{soc} level. It can be seen that the energy savings when comparing the processor and accelerator are lower than at the \ac{soc} level. It is caused by some always-on modules in \ac{soc} that introduce a power consumption offset that benefits the latter metric, although the memory subsystem is consuming more in the \ac{cgra} execution because of the simultaneous interleaved accesses.

\begin{table*}[t]
    \centering
    \noindent
    \caption{Multi-shot kernel results}
    \begin{tabular}{r | N N N N N N N N}
        \toprule
        \textbf{Kernel} & \textbf{\textit{mm 16x16}} & \textbf{\textit{mm 64x64}} & \textbf{\textit{conv2d}} & \textbf{\textit{gemm}} & \textbf{\textit{gemver}} & \textbf{\textit{gesummv}} & \textbf{\textit{2mm}} & \textbf{\textit{3mm}} \\
        \cmidrule(lr){1-9}
        \textbf{Total cycles} & 12,105 & 297,050 & 13,931 & 320,284 & 39,825 & 12,091 & 347,446 & 579,309 \\
        \textbf{Number of operations} & 7,936 & 520,192 & 65,348 & 681,000 & 144,120 & 32,670 & 603,200 & 1,071,700 \\
        \textbf{Outputs/cycle} & 2.11$\times\num{e-2}$ & 1.38$\times\num{e-2}$ & 2.58$\times\num{e-1}$ & 1.31$\times\num{e-2}$ & 3.68$\times\num{e-1}$ & 7.44$\times\num{e-3}$ & 9.21$\times\num{e-3}$ & 4.83$\times\num{e-3}$ \\
        \cmidrule(lr){1-9}
        \textbf{Performance (MOPs)} & 163.90 & 437.80 & 1,172.71 & 531.56 & 904.71 & 675.50 & 434.02 & 462.49 \\
        \textbf{CGRA consumption (mW)} & 3.99 & 7.46 & 10.11 & 9.91 & 10.36 & 8.99 & 8.66 & 8.29 \\
        \textbf{Energy efficiency (MOPs/mW)} & 41.08 & 58.66 & 115.96 & 53.62 & 87.30 & 75.16 & 50.10 & 55.80 \\
        \cmidrule(lr){1-9}
        \textbf{CPU cycles [-O3]} & 42,181 & 3,965,254 & 259,234 & 3,438,372 & 522,364 & 111,080 & 3,370,417 & 5,390,990 \\
        \textbf{CPU consumption (mW)} & 3.59 & 3.59 & 4.09 & 3.54 & 3.74 & 3.67 & 3.74 & 3.72 \\
        \cmidrule(lr){1-9}
        \textbf{Speed-up} & 3.48x & 13.35x & 18.61x & 10.74x & 13.12x & 9.19x & 9.70x & 9.31x \\
        \textbf{Energy savings (CPU vs. CGRA)} & 3.14x & 6.43x & 7.53x & 3.84x & 4.74x & 3.75x & 4.19x & 4.18x \\
        \cmidrule(lr){1-9}
        \textbf{SoC CGRA consumption (mW)} & 28.34 & 33.84 & 47.09 & 38.09 & 40.34 & 38.09 & 36.34 & 35.84 \\
        \textbf{SoC CPU consumption (mW)} & 27.34 & 27.34 & 28.09 & 26.59 & 27.59 & 28.34 & 27.59 & 27.84 \\
        \textbf{Energy savings (SoCs)} & 3.36x & 10.79x & 11.10x & 7.49x & 8.97x & 6.84x & 7.37x & 7.23x \\
        \bottomrule
    \end{tabular}
    \label{tab:multi-shot_kernels}
\end{table*}

Table \ref{tab:multi-shot_kernels} shows results for multi-shot kernels. The benchmark sizes are 16x16 and 64x64 matrixes for the \textit{mm}, 64x64 pixels for the \textit{conv2d}, and the \textit{SMALL\_DATASET} sizes defined in the Polybench collection. In contrast to the one-shot kernels, the control overhead, the configuration cycles, the execution, and the synchronization with the \ac{cpu} are considered this time in \textit{Total cycles}. The behavior of the \ac{cgra} is the same as in the previous subset of benchmarks, but in this case, the accelerator is reloaded and/or reconfigured by the processor multiple times before finishing the task. The architecture's performance in these tests provides realistic measures of how the accelerator may behave in real-life use cases. The performance metrics for \textit{mm} show that the size of the matrixes affects a lot. Every partial-kernel iteration has to process four input vectors (see Figure~\ref{fig:mapping-mm}) of 16 or 64 data, and the reloads with new parameters are not negligible with low matrix dimensions. This condition also applies to the rest of the benchmarks, although their datasets are big enough to obtain an acceptable performance. The best performance for multi-shot kernels is the \textit{conv2d} because only three iterations are needed, and the execution time of each is long enough to make control and reconfiguration times negligible. In terms of power consumption, these benchmarks obtain lower values than one-shot kernels because the \ac{cgra} is clock-gated when the \ac{cpu} is reloading the memory nodes (except for \textit{conv2d} because of its significant execution/reload proportion). For this reason, the energy savings are usually higher in these cases than for the one-shot kernels.

This extended study shows that the reconfigurable accelerator obtains similar results in both scenarios, i.e., it can be used in realistic use cases, either by accelerating the one-shot inner loop of an application or by computing whole complex applications in multiple iterations. Besides, these results show that when designing a compiler, one may have to consider the mapping strategies that can be driven to exploit a specific architecture.

\subsection{Comparison with other works}

In this section, some works are compared with the architecture proposed in this paper. First, a comparison includes qualitative and quantitative aspects, which state the context of the accelerator concerning other types of architecture and technology. Finally, some of these works are selected for performance and energy efficiency comparison using the results reported in the previous subsection.

\begin{table*}[t]
    \centering
    \noindent
    \begin{threeparttable}
    \caption{\ac{cgra} features comparison}
    \begin{tabular}{r N N N N N N N}
        \toprule
        \textbf{Metric} & \textbf{\ac{strela}} & \textbf{RipTide \cite{riptide}} & \textbf{ADRES \cite{revamp}} & \textbf{HyCube \cite{revamp}} & \textbf{Softbrain \cite{revamp}} & \textbf{\mbox{UE-CGRA} \cite{ue-cgra}} & \textbf{IPA \cite{ipa}} \\
        \cmidrule(lr){1-1}
        \cmidrule(lr){2-8}
        \textbf{Internal data synchronization} & SD & SD & TM & TM & SD & SD & TM \\
        \textbf{Irregular loops} & \cmark & \cmark & \xmark & \xmark & \xmark & \cmark & \cmark \\
        \textbf{No use of scratchpads} & \cmark & \cmark & \xmark & \xmark & \xmark & \xmark & \xmark \\
        \cmidrule(lr){1-1}
        \cmidrule(lr){2-8}
        \textbf{Control CPU} & RV32IMC & RV32EMC & - & - & - & RV32IM & OpenRISC \\
        \textbf{Total Memory Size} (KB) & 256 & 256 & 64 & 64 & 64 & 64 & 77 \\
        \textbf{\ac{cgra} Size} & 4x4 & 6x6 & 6x6 & 6x6 & 6x6 & 8x8 & 4x4 \\
        \cmidrule(lr){1-1}
        \cmidrule(lr){2-8}
        \textbf{Technology} (nm) & TSMC 65 & Intel 22 & 22 & 22 & 22 & TSMC 28 & STM 28 \\ 
        \textbf{Clock Frequency} (MHz) & 250 & 50 & 100 & 100 & 100 & 750 & 100 \\
        \cmidrule(lr){1-1}
        \cmidrule(lr){2-8}
        \textbf{\ac{soc} Area} ($mm^2$) & 2.38 & 0.50 & - & - & - & - & 0.34 \\ 
        \textbf{\ac{cgra} Area} ($mm^2$) & 0.25 & 0.25 & 0.20 & 0.165 & 0.125 & 0.28 & 0.20 \\ 
        \textbf{\ac{pe} Area} ($\mu m^2$) & 13,243 & 7,000 & - & - & - & 4,000 & 7,031 \\
        \bottomrule
    \end{tabular}
    \label{tab:feat-comparison}
    \begin{tablenotes}
    \centering
    \vspace{0.1cm}
    \item[]SD: Static Dataflow \ac{cgra}. \hspace{1cm} TM: Time-Multiplexed \ac{cgra}.
    \end{tablenotes}
    \end{threeparttable}
\end{table*}

Comparison with previous \ac{cgra} works is shown in Table \ref{tab:feat-comparison}. Some widely used \acp{cgra}, such as ADRES~\cite{adres}, HyCube~\cite{hycube}, or Softbrain~\cite{stream-dataflow}, are included using REVAMP~\cite{revamp}, a recent study of heterogeneity, although only homogeneous results are considered. These three works are evaluated for performance and energy efficiency in the paper, but the metrics are given as an average of a subset of benchmarks, even without including a control processor in the \ac{soc}. IPA~\cite{ipa} is selected because it has the same \ac{cgra} size as \ac{strela} and an ultralow-power target, using an OpenRISC control processor. UE-CGRA~\cite{ue-cgra} and RipTide~\cite{riptide} are selected because they have a static dataflow execution model and also use a \mbox{RISC-V} control processor. The first three rows classify the works into static dataflow and time-multiplexed \acp{cgra}, whether there is support for conditionals, and the ability to avoid using dedicated memories in the accelerator.  Then, the control core, memory size, and \ac{cgra} dimensions are reported. All these features match the embedded domain target, note that the maximum size of the fabric is 8x8, and the maximum memory size is 256 KB. In our case, 4x4 is enough, as many applications can be deployed with the mentioned mapping strategies. In addition, it is even more versatile for mapping one-shot kernels than bigger \acp{cgra} as RipTide (they have to split the \textit{fft} benchmark into parts to process it in multiple iterations due to heterogeneity). From this point, the differences in the technology node affect the ease of putting the proposed work in context. There is a tendency to have low clock frequencies for reducing power consumption and, in some cases, because the designs do not meet timing constraints. However, the simulations carried out for this paper proved that the energy efficiency can be even higher using 250 MHz instead of 100 MHz. Only in this work and in UE-CGRA, the frequency exceeds 100 MHz. Regarding resource utilization, previous work consistently reported less area than this work due to smaller technology nodes. However, in some cases, the differences are not that great. In IPA, the scratchpads occupy about one-third of the total area. The \ac{pe} area for IPA and RipTide is similar, but significantly larger than the UE-CGRA \ac{pe} in both cases. At \ac{soc} level, it can be seen that in other works, \acp{cgra} are a significant percentage of the total. For example, RipTide uses 50\% of the total area in \ac{cgra} because it has a \ac{noc} to interconnect their \acp{pe} that overheads it. On the contrary, in this work, \ac{cgra} area is only 10.7\%, and the memory occupies the most.

\begin{table*}[t]
    \centering
    \noindent
    \begin{threeparttable}
    \caption{\ac{cgra} performance comparison}
    \begin{tabular}{N N N N N N N N N N N }
        \toprule
        \multicolumn{2}{N }{\textbf{}} & \multicolumn{3}{c }{\textbf{Perf. (MOPs)}} & \multicolumn{3}{c }{\textbf{Power Consumption (mW)}} & \multicolumn{3}{c }{\textbf{Energy Efficiency (MOPs/mW)}} \\  
        \cmidrule(lr){3-11}
        \textbf{Work} & \textbf{Frequency (MHz)} & \textit{fft} & \textit{mm 16x16} & \textit{mm 64x64} & \textit{fft} & \textit{mm 16x16} & \textit{mm 64x64} & \textit{fft} & \textit{mm 16x16} & \textit{mm 64x64} \\
        \cmidrule(lr){1-1}
        \cmidrule(lr){2-2}
        \cmidrule(lr){3-5}
        \cmidrule(lr){6-8}
        \cmidrule(lr){9-11}
        \multicolumn{1}{ N }{\textbf{IPA\tnote{$\star$}}} & 100 & - & 65.98 & - & - & 0.49 & - & - & 134.65 & -\\
        \multicolumn{1}{ c }{\textbf{UE-CGRA\tnote{$\dagger$}}} & 750 & 625.00 & - & - & 14.01 & - & - & 44.61 & - & - \\
        \multicolumn{1}{ c }{\textbf{RipTide\tnote{$\star$}}} & 100 & 62 & - & 164 & 0.24 & - & 0.50 & 258.33 & - & 328.00 \\
        \multicolumn{1}{ c }{\textbf{\ac{strela}\tnote{$\star$}}} & 250 & 1,223.71 & 163.90 & 437.80 & 16.84 & 3.99 & 7.46 & 72.68 & 41.08 & 58.66 \\
        \bottomrule
    \end{tabular}
    \label{tab:perf-comparison}
    \begin{tablenotes}
    \centering
    \vspace{0.1cm}
    \item[$\star$] Post-synthesis results. \hspace{1cm} $\dagger$ Post-P\&R results.
    \end{tablenotes}
    \end{threeparttable}
\end{table*}

Performance and consumption results are provided in Table \ref{tab:perf-comparison}. As mentioned earlier, the works of REVAMP do not provide reproducible results and are not included in the table. IPA is suitable for performance comparison because the authors give execution time metrics for 16x16 matrix multiplication, and using the naive algorithm equation for counting operations, the value is adjusted to match the architecture-agnostic metrics. RipTide provides architecture-agnostic metrics, for itself and also UE-CGRA, which are used in the comparison. Regarding performance, \ac{strela} outperforms all previous works, even those with larger array sizes. For power consumption, this work does not achieve a good score. This is caused by the use of bigger node technology, but also because the elastic logic introduces an overhead in energy consumption (each \textit{Elastic Buffer} consumes about 80 $\mu$W when used). Finally, in the energy efficiency metrics, it can be seen that RipTide and IPA achieve an ultra-low power budget, but \ac{strela} outperforms UE-CGRA in energy efficiency as the performance is enough.

\section{Conclusion}\label{sec:conclusions}

This work proposes an elastic \ac{cgra} accelerator provided with memory streaming units, integrated into a \mbox{RISC-V}-based \ac{soc} targeting embedded applications. An enhanced microarchitecture is presented, incorporating elastic logic to support conditionals and irregular loops, demonstrating adaptability to the embedded domain's control- and data-driven applications. Additionally, the proposed mapping strategies allow exploiting the benefits of this \ac{cgra}, whether for the named one-shot kernels that fit into the architecture or multi-shot kernels that are too large to fit into it and have to be processed in several \ac{cgra} executions.

The accelerator's independent memory nodes leverage the \ac{cgra} from the address generation, enabling the execution of larger kernels in the 4x4 \ac{cgra} as only the data and control operations are performed by the \acp{pe}. Furthermore, implementing power- and clock-gating techniques optimizes the \ac{cgra} for the embedded domain, combining performance with energy consumption management.

The results show that STRELA is tailored for computing data-driven applications, performing slightly worse when tackling simple control-driven applications featuring small irregular loops. However, in scenarios involving complex control-driven kernels, the performance experiences a relative decline compared to the other cases. The best performance is 1.22 GOPs, and the best energy efficiency is 115.96 MOPs/mW. Performance improvement is reported compared to previous state-of-the-art works, outperforming them 6.71x on average. For comparison in terms of energy, a ratio of 0.60x more energy efficiency on average (or 1.67x less energy efficient) is reported, which is caused by the use of elastic logic and by the technological differences stemming from using larger technology nodes than the 22-28 nm of other works.

In conclusion, this work shows the ease of use of this elastic \ac{cgra}, proves their flexibility to execute embedded code sections that require intensive computing, provides guidelines for designing a compiler and mapper for the presented microarchitecture, and establishes scenarios where this architecture can outperform existing proposals.

\ifCLASSOPTIONcaptionsoff
  \newpage
\fi

\bibliographystyle{IEEEtran}
\bibliography{bibliography}

\begin{IEEEbiography}[{\includegraphics[width=1in,height=1.25in,clip,keepaspectratio,viewport=0 0 1786 2500]{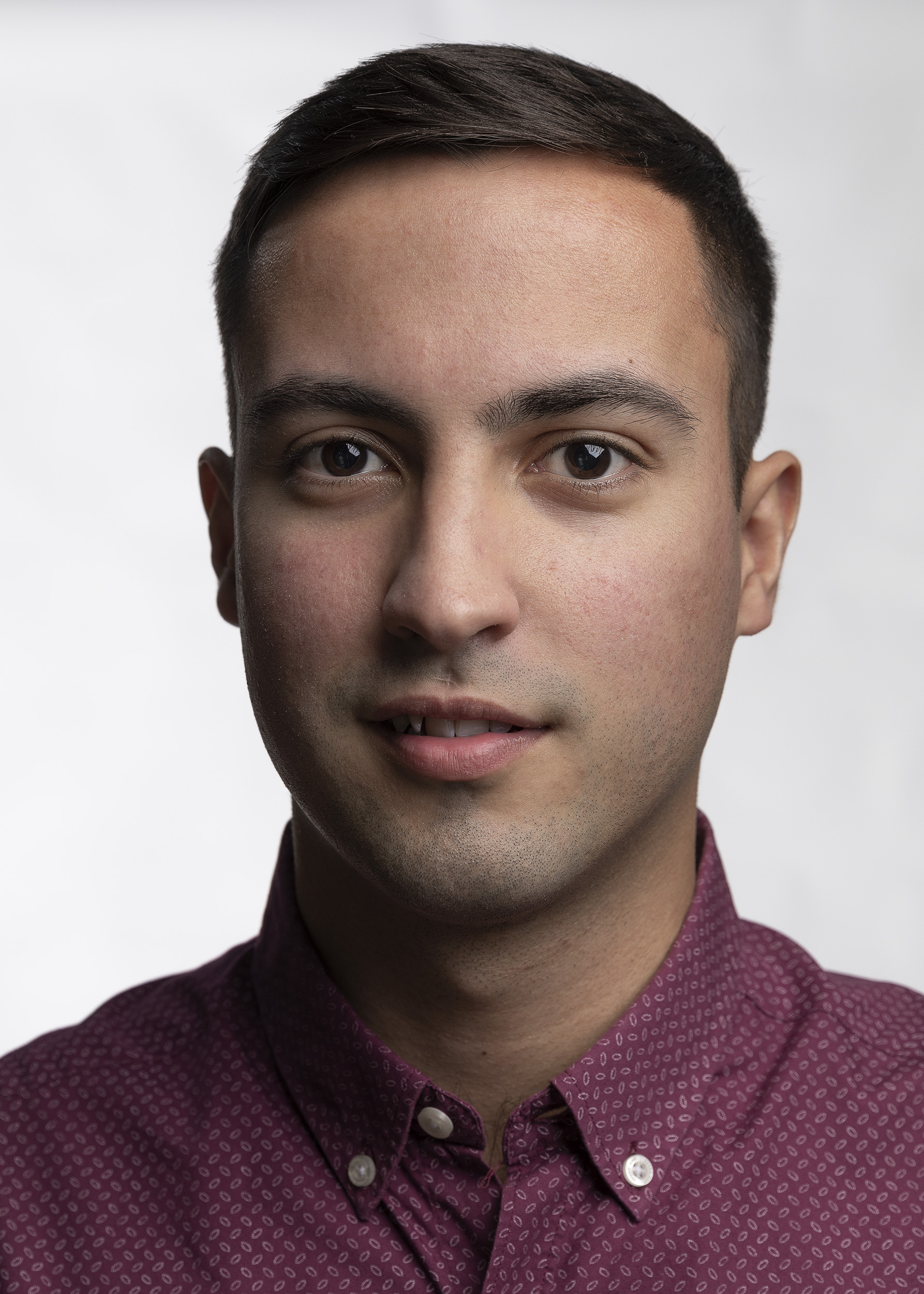}}]{Daniel V\'azquez} received his B.Sc. degree in Industrial Electronics and Automation Engineering from the Universidade de Vigo, where he graduated in 2016. He received his M.Sc. degree in Industrial Electronics from Centro de Electr\'onica Industrial (CEI) of the Universidad Polit\'ecnica de Madrid (UPM) in 2021, where he is currently a full-time researcher and Ph.D. student in Industrial Electronics. His research interests include embedded systems, reconfigurable computing, and hardware acceleration.
\end{IEEEbiography}

\begin{IEEEbiography}[{\includegraphics[width=1in,height=1.25in,clip,keepaspectratio,viewport=0 0 546 667]{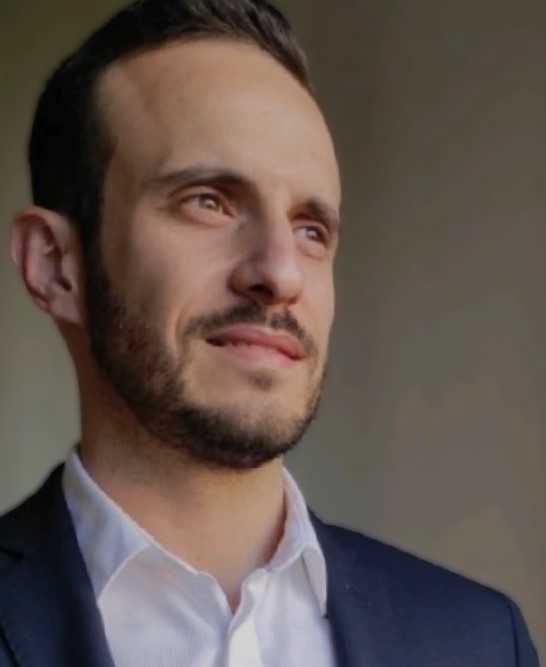}}]{Jos\'e Miranda} received his Ph.D. degree at the Microelectronic Design and Applications (DMA) research group, which belongs to Universidad Carlos III de Madrid. 
He is currently a Post-doc research assistant at the Embedded Systems Laboratory, EPFL. His research field comprises wireless sensors, embedded systems, wearable design, development and integration for safety applications, affective computing implementation into edge computing devices, and hardware acceleration. 
\end{IEEEbiography}

\begin{IEEEbiography}[{\includegraphics[width=1in,height=1.25in,clip,keepaspectratio,viewport=0 0 421 593]{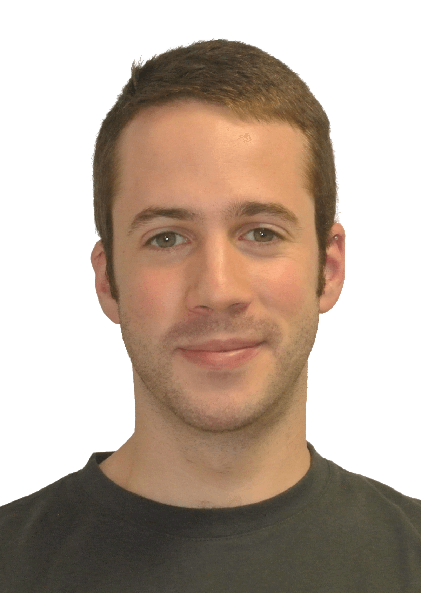}}]{Alfonso Rodr\'{i}guez} (S'15-M'21) received the BSc degree in industrial engineering, the MSc degree in industrial electronics, and the PhD degree in electrical and electronics engineering from the Universidad Polit\'{e}cnica de Madrid (UPM), Madrid, Spain, in 2012, 2014, and 2020 respectively. He is currently an Associate Professor and full-time researcher at Centro de Electr\'{o}nica Industrial, UPM. Under a HiPEAC collaboration grant, he was a visiting researcher in the Computer Engineering Group at Universit\"at Paderborn, where he worked on operating systems for reconfigurable computing. His current research interests include high-performance embedded systems, computer architecture, open hardware systems, and reconfigurable computing.
\end{IEEEbiography}

\begin{IEEEbiography}[{\includegraphics[width=1in,height=1.25in,clip,keepaspectratio,viewport=0 0 1702 2017]{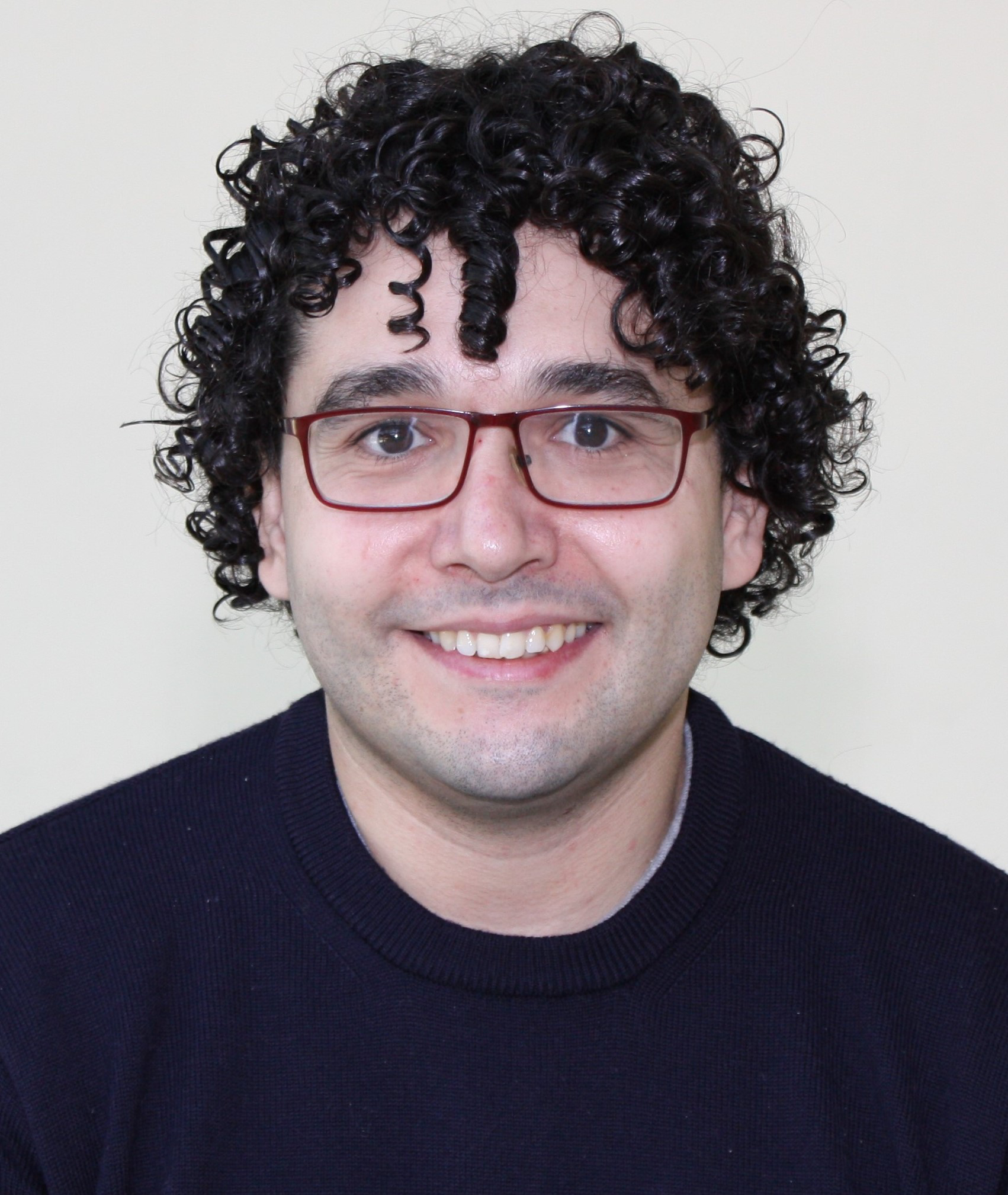}}]{Andr\'es Otero} received his M.Sc. degree in Telecommunication Engineering from the University of Vigo, where he graduated with honors in 2007. He received his Master of Research and Ph.D. degrees in Industrial Electronics from Universidad Politécnica de Madrid (UPM) in 2009 and 2014, respectively. He is currently an associate professor of electronics with the UPM, as well as a researcher in the Centro de Electrónica Industrial (CEI). His current research interests are embedded system design, reconfigurable systems on FPGAs, evolvable hardware, and embedded machine learning. During the last few years, he has been involved in different research projects in these areas, and he is the author of more than 40 papers published in international conferences and journals. He has served as the Program Committee member of different international conferences in the field of reconfigurable systems, such as SPL, ERSA, ReConFig, DASIP, and ReCoSoC.
\end{IEEEbiography}

\begin{IEEEbiography}[{\includegraphics[width=1in,height=1.25in,clip,keepaspectratio,viewport=0 0 800 800]{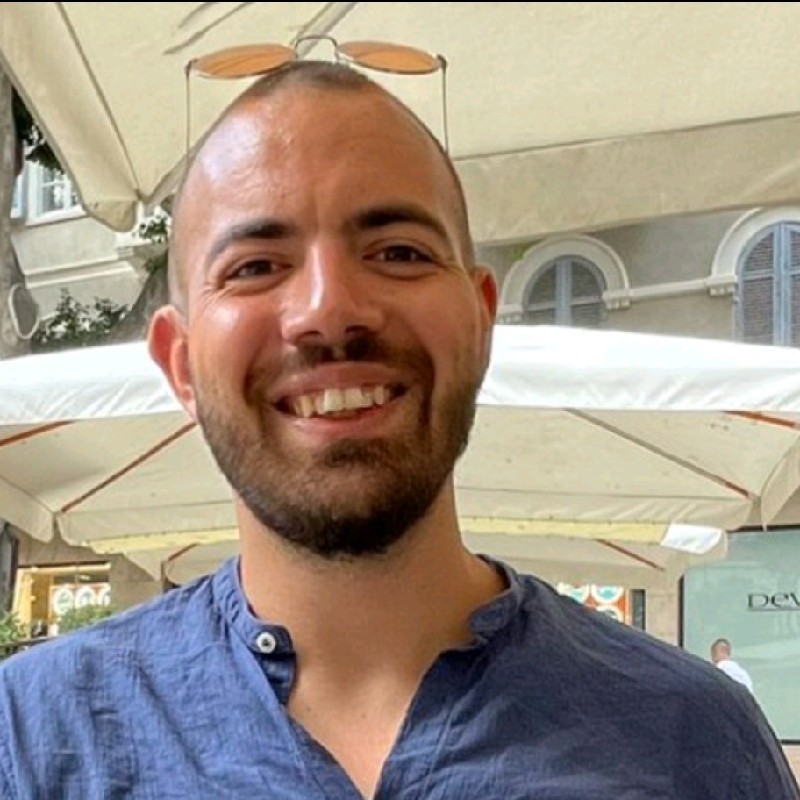}}]{Pasquale Davide Schiavone} is a PostDoc at the EPFL and Director of Engineering of the OpenHW Group. He obtained the Ph.D. title at the Integrated Systems Laboratory of ETH Zurich in the Digital Systems group in 2020 and the BSc. and MSc. from ``Politecnico di Torino'' in computer engineering in 2013 and 2016, respectively. His main activities are the RISC-V CPU design and low-power energy-efficient computer architectures for smart embedded systems and edge-computing devices.
\end{IEEEbiography}

\begin{IEEEbiography}[{\includegraphics[width=1in,height=1.25in,clip,keepaspectratio,viewport=0 0 1613 2424]{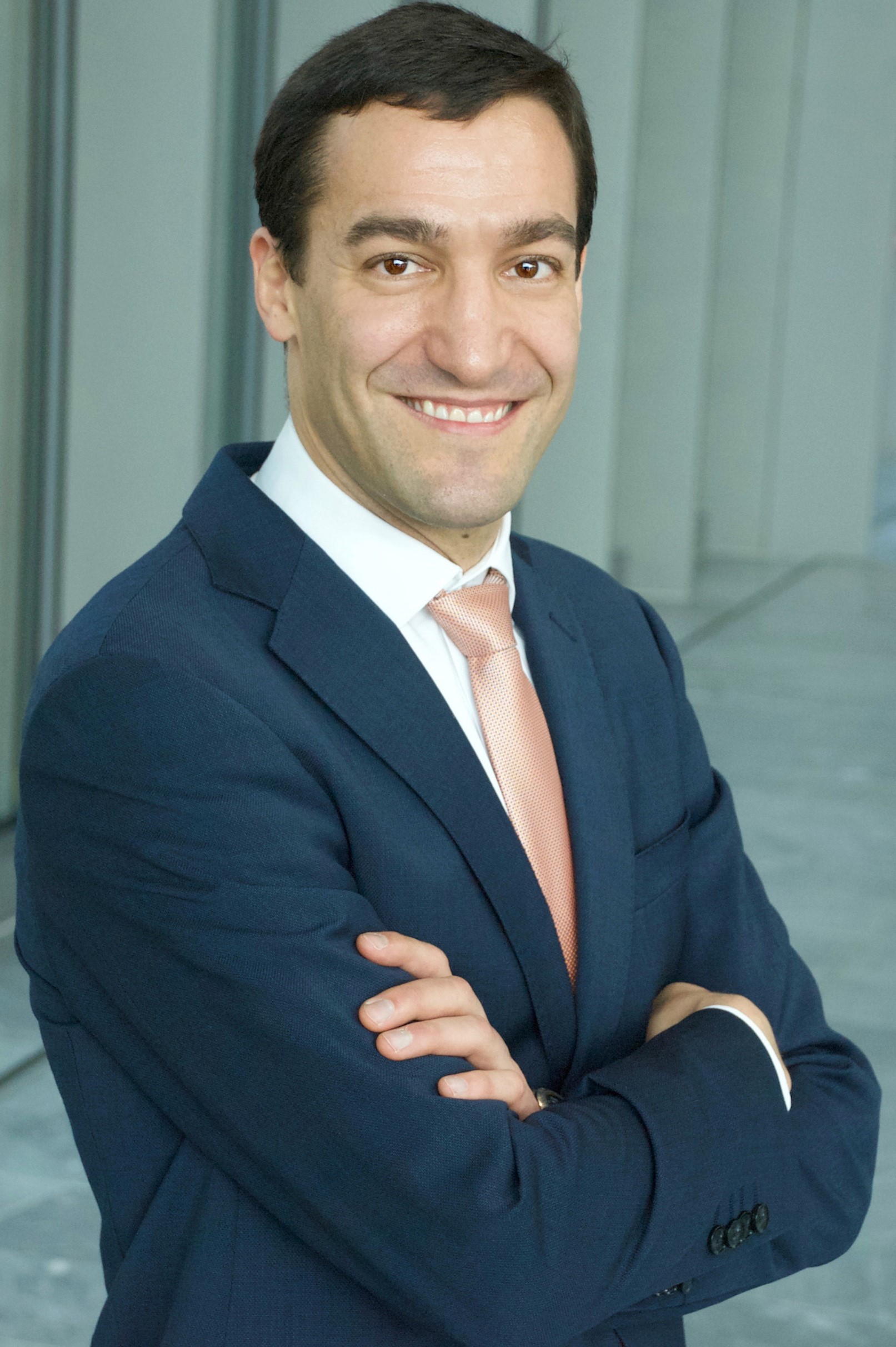}}]{David Atienza} (M'05-SM'13-F'16) received his MSc and Ph.D. degrees in Computer Science and Engineering from Complutense Univ. of Madrid (UCM), Spain, and IMEC, Belgium, in 2001 and 2005. He is a professor of Electrical and Computer Engineering, Director of the EcoCloud Center for sustainable computing, and heads the Embedded Systems Laboratory (ESL) at EPFL, Switzerland. His research interests include system-level design methodologies for high-performance multi-processor system-on-chip (MPSoC) and low-power Internet-of-Things (IoT) systems, including new 2-D/3-D thermal-aware design for MPSoCs and many-core servers, ultra-low power system architectures for wearable systems and edge AI computing, HW/SW reconfigurable systems, and memory hierarchy optimizations. He is a co-author of more than 450 publications in peer-reviewed international journals and conferences, several book chapters, and 14 patents in these fields. He has earned several best paper awards, he serves as Editor-in-Chief of IEEE TCAD (period 2022-2025)and ACM CSUR (since 2024). He was the Technical Programme Chair of IEEE/ACM DATE 2015 and General Programme Chair of IEEE/ACM DATE 2017. Dr. Atienza received the ICCAD 10-Year Retrospective Most Influential Paper Award, and the DAC Under-40 Innovators Award, among others. He is a Fellow of IEEE and a Fellow of ACM.
\end{IEEEbiography}

\vfill

\end{document}